\documentclass[prb,onecolumn]{revtex4}
\usepackage{graphicx}
\usepackage{amsmath}
\usepackage{amssymb}
\usepackage{stmaryrd}
\usepackage{color}

\definecolor{labelkey}{cmyk}{.4,.2,0,0}

\newcommand{\be}{\begin{equation}}
\newcommand{\ee}{\end{equation}}
\newcommand{\bea}{\begin{eqnarray}}
\newcommand{\eea}{\end{eqnarray}}
\newcommand{\nn}{\nonumber}

\begin{document}

\title{Large deviations for the KPZ equation from the KP equation}

\author{Pierre Le Doussal}

\affiliation{%{$^2$} 
Laboratoire de Physique de l'\'Ecole Normale Sup\'erieure, PSL University, CNRS, Sorbonne Universit\'es, 24 rue Lhomond, 75231 Paris, France.}

\begin{abstract}
Recently, Quastel and Remenik \cite{QRKP} [arXiv:1908.10353] found a remarkable relation between some solutions
of the finite time Kardar-Parisi-Zhang (KPZ) equation and the Kadomtsev-Petviashvili (KP) equation. 
Using this relation we obtain the large deviations at large time and at short time for the KPZ equation
with droplet initial conditions, and at short time with half-Brownian initial conditions. 
It is consistent with previous results and allows to obtain sub-leading corrections, as well as
results at intermediate time.
In addition, we find that the appropriate generating function associated to the 
full Brownian initial condition also satisfies the KP equation. Finally, 
generating functions for some linear statistics of the Airy point process
are also found to satisfy the KP property, and consequences are discussed.
\end{abstract}
\maketitle

\tableofcontents

%\begin{figure}[htb]
%\includegraphics[width=.23\textwidth]{modelAnderson.pdf}
%\includegraphics[width=.23\textwidth]{modelNSS.pdf}
%\caption{Scheme of the geometrical arrangements of the a) Anderson model, and the b) NSS model.
%Blue dots are site with random disorder energies  and red dots sites on the leads,
%without disorder. Thin lines correspond to the bulk hopping strength equal 1, while thick
%lines along the edge to $t$. }
%\label{fig0}
%\end{figure}

\section{Introduction}

The Kardar-Parisi-Zhang equation \cite{KPZ} in one dimension is a continuum model for the stochastic growth of the height field $h(x,t)$, $x \in \mathbb{R}$, as a function of time $t$, of an interface between two phases in a two dimensional geometry. It reads 
\be
 \partial_t h = \partial_x^2 h + (\partial_x h)^2 + \sqrt{2} \, \xi(x,t) 
\ee
in the units chosen here, where $\xi(x,t)$ is a unit Gaussian space-time white noise. It maps to the equilibrium statistical mechanics problem of a directed polymer in a $d=1+1$ random potential \cite{kardareplica}, of partition function $Z(x,t)=e^{h(x,t)}$, which satisfies the stochastic heat equation (SHE)
\be
\partial_t Z = \partial_x^2 Z + \sqrt{2} \, \xi(x,t) Z 
\ee
defined here with the Ito prescription. Three initial conditions (IC) have been much studied (i) the flat IC 
$Z(x,t=0)=1$, (ii) the droplet IC, $Z(x,t=0)=\delta(x)$, and (iii) the Brownian IC, $Z(x,t=0)= e^{B_L(x)+w_L x} \theta(-x)
+ e^{B_R(x)-w_R x} \theta(x)$ where $B_{L,R}(x)$ are two unit half-sided Brownian motions with $B_{L,R}(0)=0$.
The case $w_{L,R} \to 0$ is of special interest as it corresponds to the stationary IC. \\

The KPZ equation is at the center of a vast universality class, the
KPZ class, which, in one dimension (to which we restrict here), contains a number of solvable discrete models for e.g. growth \cite{png,spohn2000,ferrari1}, particle transport \cite{spohnTASEP,Airy1TASEP}, or polymers \cite{Johansson2000}. Exact solutions have also been obtained for the one-point cumulative distribution function (CDF) of the height at arbitrary time for the KPZ equation, for the three aforementioned special initial conditions
\cite{we,dotsenko,spohnKPZEdge,corwinDP,sineG,we-flat,we-flatlong,SasamotoHalfBrownReplica,QuastelAi2BM,BCF,SasamotoStationary,SasamotoStationary2,BCFV}. These results exhibit universal convergence at large time, upon scaling $h$ with $t^{1/3}$ and
space $x$ with $t^{2/3}$, to Tracy-Widom (TW) type distributions \cite{TW1994Airy}, the precise type depending on the class of initial conditions, specifically the GOE-TW distribution for flat IC, the GUE-TW distribution for droplet IC and the Baik-Rains distribution \cite{png} for stationary KPZ.\\

Recently, a very detailed characterization of the universal KPZ fixed point, which governs the 
infinite time limit of all models in the KPZ class,
has been obtained from the large time asymptotics of the TASEP model, for essentially arbitrary deterministic initial condition
\cite{KPZFixedPoint,quastel2019flat}. 
The single-time, multi-point CDF of the (properly scaled) height field $h(x,t)$ was expressed as a Fredholm determinant (FD) with a Airy type kernel, quite complicated and non explicit for general IC (formally constructed from a 
Brownian scattering operator) but which simplifies into explicit forms for a number of cases. In parallel, asymptotic results were also obtained for the TASEP and the KPZ class on a finite size periodic ring
\cite{PeriodicKPZProlhac2019,BaikLiuPeriodicMultiPoint,ProlhacTASEP2015,LiuMultiTime}.\\

More recently, from the general FD form, Quastel and Remenik showed \cite{QRKP}
that the
CDF of the (properly scaled) height field can be related to solutions of a well known equation in the
theory of integrable systems, the Kadomtsev-Petviashvili (KP) equation (for the one-point CDF) and the
KP matrix equation for the multi-point CDF. This remarkable result, which holds for
the KPZ fixed point, i.e. for the infinite time limit of the typical fluctuations of the height field,
was termed "completely unexpected". The appearance of KP-like solitons in the description of the KPZ fixed point was also pointed out in \cite{PeriodicKPZProlhac2019}.\\

Even more surprising, it was noted that the finite-time solution of the KPZ equation itself
can be related (for arbitrary time) to the KP equation. This was obtained in Ref. \cite{QRKP} for the droplet and half-Brownian initial conditions.
More precisely, let us define the following generating function for the
KPZ height at one point $x$ and time $t$, equivalently the Laplace transform of the probability distribution function (PDF) of $Z(x,t)$,
\be \label{gen0} 
G(x,t,r) = \big\langle \exp(- e^{ h(x,t) + \frac{t}{12} - r}) \big\rangle 
= \big\langle \exp(- Z(x,t) e^{\frac{t}{12} - r}) \big\rangle
\ee 
where $\langle \cdots \rangle$ denotes the
average w.r.t. the KPZ noise $\xi(x,t)$. It was shown a while back, using e.g. the replica Bethe ansatz,
that this generating function 
can be calculated exactly for all $t$, as a Fredholm determinant (FD), for the droplet IC \cite{we,dotsenko,spohnKPZEdge,corwinDP,sineG}
and for the half-Brownian IC \cite{SasamotoHalfBrownReplica,QuastelAi2BM}
, that is (on $x>0$) $Z(x,t=0)= e^{B(x)-w x} \theta(x)$. 
It was obtained in Ref. \cite{QRKP}, using the FD expressions, 
that the following function $\phi(r,x,t)$ of three variables
\be \label{defphi} 
\phi(x,t,r) = \partial_r^2 \log G(x,t,r) 
\ee
obeys the KP equation
\be \label{KP1} 
\partial_t \phi + \phi \, \partial_r \phi + \frac{1}{12} \partial_r^3 \phi + \partial_r^{-1} \partial_x^2 \phi
= 0
\ee
The question of the initial condition is somewhat subtle and is discussed below. 
Note that for any fixed $x,t$, $G(x,t,r)$ increases from $0$ (for $r=-\infty$)
to $1$ (for $r=+\infty$), hence $\log G(x,t,r) \leq 0$ is increasing, i.e. $\partial_r \log G(x,t,r)$ is positive and 
$G(x,t,r)$ and its derivatives w.r.t. $r$ vanish at $r=+\infty$.\\

The unexpected result \eqref{defphi}-\eqref{KP1} opens many questions, and we wish to address some of them here. 
We also wish to open the new toolbox that it provides. We first show that one can easily recover recent results about the large deviations for the KPZ equation, directly from the KP equation, and obtain a bit more. Large deviations mean rare fluctuations, away from the well-studied typical fluctuations $H \sim t^{1/3}$. There are two limits of interest, large time $t \gg 1$ and short time $t \ll 1$. It was shown that for droplet IC, 
the PDF of the (shifted) one point height $H=h(x,0)+\frac{t}{12}$, takes at large time, and
in the scaling region $H \sim t$, the large deviation form \cite{LargeDevUs}
\be
P(H,t) \sim \exp(- t^2 \Phi_-(\frac{H}{t}) ) \label{LDLT1} 
\ee
This holds for the left large deviation tail $H/t<0$ (similar results hold for the right tail, with a different rate, not
addressed here). The exact rate function $\Phi_-(z)$ was obtained by {\it four} different, and non-trivial methods involving: (i) the WKB limit of a non-local Painleve equation \cite{sasorov2017large} (ii) the free energy of a Coulomb gas perturbed at the edge \cite{JointLetter}
(iii) a variational formula using the stochastic Airy operator \cite{SAOTsai2018},
(iv) a summation of the {\it short time} expansion using cumulants \cite{ShortTimeSystematic} 
(see related rigorous results
for the tails in \cite{CorwinTails}). In Ref. \cite{largedev} 
it was found how the four methods can be related and extended to treat a broader class of problems, involving linear statistics of the eigenvalues at the edge of the $\beta$-random matrix ensembles
(described by the Airy$_2$ point process). Here we show that \eqref{LDLT1}, 
together with the exact expression for $\Phi_-(z)$, arise quite naturally from the analysis of the KP equation, henceforth providing a fifth method. In addition, we extract the subleading corrections at large time.\\

At short time $t \ll 1$, it was shown that the large deviations occur in the regime $H \sim O(1)$
(while the typical fluctuations are $H \sim t^{1/4}$) and take, quite generically, the form 
\be
P(H,t) \sim \exp(- \frac{\Phi(H)}{\sqrt{t}} )
\ee 
The rate function $\Phi(H)$ was (i) calculated from exact solutions for droplet IC, Brownian IC 
and for some half-space cases \cite{le2016exact,krajenbrink2017exact,KrajLedou2018,krajenbrink2018large,AlexKrajPHD},
(ii) related to solutions of (saddle point) differential equations using
the weak noise theory, allowing to extract the exact asymptotics of $\Phi(H)$ at large $|H|$
for a variety of IC \cite{Korshunov,Baruch,MeersonParabola,janas2016dynamical,meerson2017randomic,
Meerson_Landau,Meerson_flatST,asida2019large,meerson2018large,smith2018finite,smith2019time}.
Both methods were found to be consistent, and the results were also tested in very high precision
numerical simulations \cite{NumericsHartmann,HartmannHalfSpace,HartmannOptimalpaths}.
The first method proceeds by first showing that at small time, for e.g. the droplet IC one has,
for any $z>0$
\be \label{Ppsi} 
\big\langle \exp(- \frac{z e^H}{\sqrt{t}}) \big\rangle  \sim \exp(- \frac{\Psi(z)}{\sqrt{t}} ) \quad , \quad 
\Psi(z) = \frac{-1}{\sqrt{4 \pi}} {\rm Li}_{5/2}(-z)
\ee 
with $H=h(0,t)+\frac{1}{2} \log t$. The rate function $\Phi(H)$ is then obtained via a (quite subtle)
Legendre transform. In Ref. \cite{ShortTimeSystematic}, we further calculated, for droplet IC, the subleading terms in
the exponential in the r.h.s. of \eqref{Ppsi}, which takes the form of the series $\frac{1}{\sqrt{t}} \Psi(z) + \sum_{p \geq 1} t^{\frac{p-1}{2}} \Psi_p(z)$, up to a very high order, $O(t^3)$.\\

To address the small time large deviations, we first write the property \eqref{defphi}-\eqref{KP1}
in terms of equations satisfied by the cumulants of $Z(x,t)$. Using the known expressions for the first lowest cumulants, we check that these equations are indeed satisfied for the droplet and the half-Brownian initial
conditions (but not for the flat IC, which thus does not seem to be simply related to KP). One finds 
that, at short time, the non-linear term in the KP equation enters only perturbatively.
This allows to determine iteratively the subleading terms in the r.h.s. of \eqref{Ppsi} 
in terms of only the leading one, $\Psi(z)$. This procedure very efficiently
recovers the systematic expansion obtained in Ref. \cite{ShortTimeSystematic}, and 
allows to go beyond. This in itself, provides a strong test of the KP property. 
However, we find that for the droplet IC the initial data problem is subtle, i.e
the leading term, $\Psi(z)$, remains undetermined. Specifying this large deviation rate function
 is equivalent to
specifying the amplitudes of the $\sim 1/\sqrt{t}$ leading short time behavior 
for each cumulant of $Z(0,t)$. This thus appears as the initial data one must input in the KP equation. 
This subtlety  is not so surprising, since the droplet IC, $Z(x,0)=\delta(x)$,
needs some regularization, see below.\\

In Section \ref{subsec:cumulants}, we provide a bridge between the short time and the 
large time large deviations. This is achieved through a summation of cumulants, initiated
in Ref. \cite{ShortTimeSystematic}, and that we push here, thanks to the KP equation,
to the next subdominant order. We show that it is equivalent to a semi-classical expansion,
which takes the form of a perturbative expansion in the third derivative in the KP equation, whose
leading order amounts to solve the Burgers equation. 

The generating function \eqref{gen0} of the KPZ equation with droplet initial condition
can be put in the form $G(x,t,r)=\hat G(t,r + \frac{x^2}{4t})$, where $\hat G(t,r)$
satisfies a reduced version of the KP equation (known as cylindrical KdV equation). 
We show that the KPZ equation with droplet IC is only one particular solution of a 
more general class of solutions, which encode for some linear statistics of the Airy$_2$ point process
(denoted $a_j$) 
\be \label{hatGgen} 
\hat G(t,r)= \log \mathbb{E}_{\rm Ai}[ \exp\big(  \sum_{j=1}^\infty f(t^{1/3} a_j - r) \big)]
\ee
This generating function was studied in Refs. \cite{ShortTimeSystematic,largedev} (see definitions
in Eqs (30-32) there), and $f(x)$ is a fairly general function, the special choice $f(x)= - \log(1+e^x)$
corresponds to the KPZ equation. We show here that for any $g(x)$ (where it exists),
$\partial_r^2 \hat G(t,r)$ satisfies the reduced KP equation, see Remark 2. below,
and the Appendix \ref{app:fd} for the general analysis of a family of FD which satisfy the KP equation.
This allows for a semi-classical expansion of the linear statistics of the Airy$_2$ point process
using the KP equation, discussed in Appendix \ref{app:g}. \\

Next, we consider the half-Brownian initial condition. There, using the KP equation,
we obtain in the short time limit $t\ll 1$,  
with $\tilde x=x/\sqrt{t}$ and $\tilde w=w \sqrt{t}$ being kept fixed
\be \label{Ppsi2} 
G(x,t,r) \simeq
\big\langle \exp(- \frac{z e^H}{\sqrt{t}}) \big\rangle  \sim \exp( - \frac{\Psi_{\tilde w}(\tilde x=\frac{x}{\sqrt{t}}
,\sqrt{t} z) )}{\sqrt{t}}
\ee 
with $H=h(x,t) + \frac{1}{2} \log t$, together with an explicit expression for $\Psi_{\tilde w}(z,\tilde x)$, see
\eqref{explicit2} and \eqref{alternate}. Further taking the
limit $\tilde w \to +\infty$ we finally obtain the result \eqref{Ppsi} for the droplet IC. Hence the 
half-Brownian solution to the KP equation, which is well defined at $t=0$,
can be used to regularize the droplet solution at small time.\\

Finally, comparing the equations that the cumulants of $Z(x,t)$ must obey so that KP holds,
and the known expressions for these cumulants, e.g. from the replica Bethe ansatz, 
we identify the mechanism of solvability, see Section \ref{sec:half} and Appendix \ref{app:bethe}
. It then implies that any IC which has a
"decoupled" overlap with the Bethe eigenstates will similarly obey the KP equation.
This is the case for the droplet and the half-Brownian IC, but since this is also
the case for the full Brownian initial condition, we conclude that the full Brownian IC
does also satisfy KP. More precisely, using an appropriately modified generating function $\tilde G$,
$\phi = \partial_r^2 \tilde G$ must satisfy KP. We check explicitly this conjecture
by comparing with the known small time large deviation rate function for the Brownian IC obtained in 
Ref. \cite{krajenbrink2017exact}. \\

Note that the original theorem for the CDF of the KPZ fixed point obeying KP was shown a priori only for
deterministic initial conditions. The fact that the KPZ equation with random IC (Brownian and half-Brownian)
also obeys KP is thus a quite interesting development \cite{footnoteQR}. \\

{\it Note added}. In a recent preprint \cite{CafassoTail}, simultaneous with the first version of this work,
Cafasso and Claeys obtain yet another derivation of the KPZ large deviation left tail using Riemann-Hilbert (RH) methods. These methods allow for a rigorous proof of the asymptotics. 
In Remark 4 below we compare with our results. Their formula also applies to
intermediate times. We show in Section \ref{sec:related} that intermediate time results can also be obtained from the KP equation \cite{footnoteC2}. 

\section{Droplet initial condition}

%\subsection{Space dependent KP equation} 
%
%Let us consider droplet initial conditions, which are best defined for the SHE, as
%$Z(x,t)=\delta(x)$. It was shown in 2010 that the following generating function for the
%KPZ height at one point $x$ and time $t$, equivalently the Laplace transform of the PDF of $Z(x,t)$,
%\be \label{gen0} 
%G(x,t,r) = \big\langle \exp(- e^{ h(x,t) + \frac{t}{12} - r}) \big\rangle 
%= \big\langle \exp(- Z(x,t) e^{\frac{t}{12} - r}) \big\rangle
%\ee 
%can be calculated as a Fredholm determinant (FD). Here $\langle \cdots \rangle$ denotes the
%average w.r.t. the KPZ noise $\xi(x,t)$. It was obtained in Ref. \cite{QRKP}, using the
%FD expression, that the following function $\phi(r,x,t)$ of three variables
%\be
%\phi(r,x,t) = \partial_r^2 \log G(x,t,r) 
%\ee
%obeys the KP equation
%\be \label{KP1} 
%\partial_t \phi + \phi \, \partial_r \phi + \frac{1}{12} \partial_r^3 \phi + \partial_r^{-1} \partial_x^2 \phi
%= 0
%\ee
%The question of the initial condition will be discussed below. Note that for any fixed $x,t$, $G(x,t,r)$ increases from $0$ (for $r=-\infty$)
%to $1$ (for $r=+\infty$), and that $\partial_r \log G(x,t,r)$ is positive and vanishes in both limits.\\

\subsection{Space-independent, reduced KP equation} 

We start with the droplet initial condition $Z(x,t=0)=\delta(x)$. We first use the statistical tilt symmetry to 
eliminate the spatial variable $x$. For the droplet IC it is well known that
$h(x,t) \equiv h(0,t) - \frac{x^2}{4 t}$ where $\equiv$ means identity in law of the one point distributions. Hence
one can write
\be \label{GGtilde} 
G(x,t,r) = \big\langle \exp(- e^{ h(0,t) - \frac{x^2}{4 t} + \frac{t}{12} - r}) \big\rangle = \hat G(t, r +\frac{x^2}{4 t})
\quad , \quad \phi(x,t,r) = \partial_r^2 G(x,t,r)= \psi(t, r +\frac{x^2}{4 t})
\ee
where we now denote 
\be \label{defpsi} 
\hat G(t,r) = \langle \exp(- e^{ h(0,t) + \frac{t}{12} - r}) \rangle \quad , \quad \psi(t,r)= \partial_r^2 \log \hat G(t, r)
\ee
Here $\hat G$ is the standard generating function for the height at $x=0$. Injecting the form \eqref{GGtilde} into
\eqref{KP1} one obtains, after some cancellations, a reduced KP equation
\be \label{KP2} 
 \partial_t \psi +  \psi \partial_r \psi + \frac{1}{12} \partial_r^3 \psi  +  \frac{1}{2 t}  \psi =0
\ee 
Note that this equation can be integrated once, with $\psi(t,r) = \partial_r \hat \psi(t,r)$, i.e. it can 
also be written for the function $\hat \psi(t,r)= \partial_r \log \hat G(t,r)$, as
\be \label{KP22} 
 \partial_t \hat \psi + \frac{1}{2} (\partial_r \hat \psi)^2 + \frac{1}{12} \partial_r^3 \hat \psi  +  \frac{1}{2 t}  \hat \psi =0
\ee 
where the integration constant must vanish since $\hat \psi=\partial_r \log \hat G$ 
vanishes at $r \to + \infty$.\\

{\bf Remark 1.} 
Eq. \eqref{KP2} is also called the cylindrical Korteweg-de Vries (KdV) equation
(up to a rescaling of coefficients). Upon the change of variable (with $b>0$)
\be
\psi(t,r) = \frac{r}{2 t} + \frac{B}{2 t} u(y,\tau) \quad , \quad r= - \frac{y}{\tau} \quad , \quad t = \frac{b}{\tau^2} 
\ee 
it is transformed into the standard KdV equation \cite{CafassoPC} 
\be \label{KdV} 
\partial_\tau u + \frac{b}{6} \partial_y^3 u + B u \partial_y u = 0
\ee
The canonical form is obtained for $b=6$ and $B=-6$, and the form $b=\frac{1}{2}$, $B=1$
arises in the description of the KPZ fixed point for flat IC \cite{QRKP}. Note that in 
the forthcoming paper \cite{CafassoKdV}, the KdV equation \eqref{KdV} is 
derived for the general case of $\hat G(t,r)$ defined as in \eqref{hatGgen} 
using the Riemann-Hilbert setting proposed in \cite{CafassoTail}.

\subsection{Checks and moment expansion} 

We can now perform a few checks. 
The function $\psi(t,r)$, from its definition \eqref{defpsi}, admits an expansion in powers of $e^{-r}$ (e.g. for large positive $r$) whose 
coefficients are related to the cumulants $Z_n(t) = \langle Z(0,t)^n \rangle_c$ of the solution
to the SHE, as 
\be \label{exppsi} 
\psi(t,r) =   \sum_{n \geq 1} \frac{(-1)^n n^2}{n!} Z_n(t) e^{\frac{n t}{12} -n r} 
\ee 
using that $Z(x,t)=e^{h(x,t)}$.
Inserting into \eqref{KP2}, we find that the reduced KP equation implies that the 
cumulants $Z_n(t)$, $n \geq 1$, must satisfy the
following recursion relation
\bea \label{rec1} 
- \partial_t Z_n(t) 
+ \frac{n^3-n}{12} Z_n(t) 
- \frac{1}{2 t} Z_n(t)   = - \frac{(n-1)!}{2} \sum_{n_1+n_2=n, n_1,n_2 \geq 1} \frac{ 
n_1^2 n_2^2}{n_1! n_2!} 
Z_{n_1}(t) Z_{n_2}(t) 
\eea 
Before performing some simple checks that the known expressions for the $Z_n(t)$
do indeed satisfy this equation, we first ask whether this recursion determines 
the $Z_n(t)$. The answer is that this recursion determines iteratively the $Z_n(t)$,
except that at each level $n \geq 1$ there is an unknown integration constant $c_n$,
since the solution to the homogeneous part of \eqref{rec1} is 
$Z^{\rm hom}_n(t)=\frac{c_n}{\sqrt{t}} e^{\frac{1}{12} (n^3-n) t}$. 
Let us examine the two lowest orders from \eqref{rec1}
\bea
&& - \partial_t Z_1(t)  - \frac{1}{2 t} Z_1(t)   = 0 \\
&& - \partial_t Z_2(t) + \frac{1}{2} Z_2(t) - \frac{1}{2 t} Z_2(t)   = -  \frac{1}{2} Z_{1}(t)^2
\eea
The first equation gives $Z_1(t)=c_1/\sqrt{t}$. We known that, upon averaging the SHE (with the Ito convention), the first moment $\langle Z(x,t) \rangle$ satisfies the standard heat equation. Hence, for droplet IC
one has $\langle Z(x,t) \rangle = \frac{1}{\sqrt{4 \pi t}} e^{- \frac{x^2}{4 t}}$, i.e. the free diffusion kernel,
and one must have $Z_1(t)=\langle Z(0,t) \rangle= \frac{1}{\sqrt{4 \pi t}}$, which determines $c_1=1/\sqrt{4 \pi}$. 
The general solution to the second equation is then
\be \label{Z2res1}
Z_2(t)=e^{t/2} ( \frac{c_2}{\sqrt{t}} + \frac{1}{4 \sqrt{2 \pi t}} {\rm Erf}(\sqrt{\frac{t}{2}}))
\ee 
On the other hand, the result of the Bethe ansatz calculation, i.e. Eq. (11) in Ref. \cite{we}, 
is 
%\be
%\frac{Z_2(t)}{Z_1(t)^2} = \sqrt{2 \pi} (t/4)^{1/2} e^{t/2} (1+ {\rm Erf}(\sqrt{\frac{t}{2}}))
%\ee
%Hence
\be \label{Z2res2} 
Z_2(t)=\frac{1}{4} \frac{1}{ \sqrt{2 \pi t}} e^{t/2} (1+ {\rm Erf}(\sqrt{\frac{t}{2}}))
\ee
The two formula \eqref{Z2res1} and \eqref{Z2res2} are indeed consistent, provided we
choose the integration constant $c_2=\frac{1}{4 \sqrt{2 \pi}}$. It is easy, but tedious, to check 
from the Bethe ansatz results, that a similar pattern holds for higher cumulants
(see Appendix \ref{app:bethe} 
for related checks). Hence we conclude that the coefficients $c_n$
play the role of the initial data. Equivalently, one can characterize the initial data
by specifying the small time $t$ limit as $Z_n(t) \simeq \frac{c'_n}{\sqrt{t}}$
(with, as we checked here $c'_n=c_n$ for $n=1,2$). 

\subsection{Short time large deviation expansion} 

It turns out that the coefficients $c'_n$ are known, they were obtained in the large deviation analysis of the
KPZ equation at short time in Ref. \cite{le2016exact}. It was shown there that as $t \to 0$
\be \label{Psi} 
\log \hat G(t,r) \simeq - \frac{1}{\sqrt{t}} \Psi(z=e^{-r})  \quad , \quad \Psi(z)=\frac{-1}{\sqrt{4 \pi}} {\rm Li}_{5/2}(-z) 
\ee
Hence the short time behavior of $\psi(t,r)$ in \eqref{exppsi} must be as $\sim \frac{1}{\sqrt{t}}$, uniformly in $r$, with
\bea \label{stld} 
&& \psi(t,r)  \simeq - \frac{1}{\sqrt{t}} \, \partial_r^2 \Psi(z=e^{-r})
= \frac{1}{\sqrt{4 \pi t}} {\rm Li}_{1/2}(-e^{-r}) = -\frac{e^{-r}}{
   \sqrt{4 \pi t }} + \frac{e^{-2
   r}}{2 \sqrt{2 \pi t}} + O(e^{-3 r}) 
\eea
This is consistent with the terms $n=1,2$ in \eqref{exppsi} and the values $c'_1=c_1=1/\sqrt{4 \pi}$,
$c'_2=c_2=\frac{1}{4 \sqrt{2 \pi}}$ obtained above (since at small $t$ one can set the factor 
$e^{\frac{n t}{12}}$ to unity to the leading order). The equation \eqref{stld} determines all the $c_n'$,
hence the initial data.\\

Let us backtrack one step and ask whether one can obtain the full small time expansion 
from the KP equation ? In Ref. \cite{ShortTimeSystematic} it was shown that \eqref{Psi}
was just the leading order of a systematic short-time expansion in powers of $\sqrt{t}$. 
Hence we will look for the form 
\be \label{serpsi} 
\psi(t,r)= \frac{1}{\sqrt{t}} \, \partial_r \bigg( p_0(r) + \sum_{m \geq 1} p_m(r) t^{\frac{m}{2}} \bigg) 
\ee
and insert it in the reduced KP equation \eqref{KP2}. It gives a recursion (see Appendix \ref{app:shorttime}) 
which can be solved in a hierarchical way 
\bea \label{hier} 
&& p_1(r) = - p'_0(r)^2 \\
&& p_2(r)  = - p'_0(r) p'_1(r) - \frac{1}{12} p_0'''(r) = \partial_r (\frac{2}{3} p'_0(r)^3 - \frac{1}{12} p_0''(r))
\eea
and so on (here and below we use indifferently the notations $\partial_r p(r) \equiv p'(r)$).
We note that $p_0(r)$ is left undetermined, but all the $p_i(r)$ with
$i \geq 1$ are obtained from $p_0(r)$. We will thus consider $p_0(r)$ as
an "initial condition" and set it equal to the known result
\be \label{init} 
p'_0(r)=  \frac{1}{\sqrt{4 \pi}} {\rm Li}_{1/2}(-e^{-r})
\ee 

We can now compare with the result from Ref. \cite{ShortTimeSystematic}, obtained through a quite complicated
calculation directly expanding the FD. The relevant formula there are (5), (30), (58), (61), (62). We check that 
\be
\psi(t,r)= \partial_r^2 q_{t,\beta=1}(\sigma) \quad , \quad \sigma=-e^{-r} \quad , \quad \sigma \partial_\sigma = - \partial_r 
\ee 
with $q_{t,\beta=1}$ there equals $\log \hat G$ here, and 
the following functions were introduced there
\be
{\cal L}_i(\sigma) = \frac{1}{\sqrt{4 \pi}} {\rm Li}_{\frac{3}{2}-i}(\sigma) \quad , \quad \sigma \partial_\sigma {\cal L}_i = {\cal L}_{i+1} 
\ee 
Let us check here the first two terms. One has $p'_0(r) = {\cal L}_1(-e^{-r})$
hence \eqref{serpsi} and \eqref{hier} lead to 
\be
\psi(t,r)=  \frac{1}{\sqrt{t}} \left( {\cal L}_1 - \sqrt{t} \partial_r {\cal L}_1^2 + t \partial_r^2 
(\frac{2}{3} {\cal L}_1^3 - \frac{1}{12} \partial_r {\cal L}_1) + O(t^{3/2}) \right)
\ee 
where ${\cal L}_i \equiv {\cal L}_i(-e^{-r})$. Using that $- \partial_r {\cal L}_i = {\cal L}_{i+1}$ and comparing
with (58) and (62) in Ref. \cite{ShortTimeSystematic} we see that it agrees. In the Appendix
\ref{app:shorttime}
we have checked terms to a much higher order, we recover all terms displayed in Ref. \cite{ShortTimeSystematic}
and show one order more, i.e. the term $O(t^{7/2})$. The present method is clearly much faster.
 \\

The conclusion of this subsection is that the KP equation allows to recover easily the full systematic short time expansion for the KPZ generating function with droplet IC, obtained in Ref. \cite{ShortTimeSystematic} from the FD, 
provided it is given as an input, i.e. initial data, the leading term \eqref{init} for the large deviations
which encodes the limits $c'_n = \lim_{t \to 0} t^{1/2} Z_n(t)$.\\

{\bf Remark 2.} It is interesting to note that in Ref. \cite{ShortTimeSystematic} we have considered a more
general problem, the evaluation of the following linear statistics over the Airy$_2$ point process
$a_i$ (see definitions and Eq. (30) and (32) there) 
\be \label{remark1} 
q_{t,\beta}(\sigma)=\log \mathbb{E}_{\rm Ai} 
\bigg[ \exp\left( \beta \sum_{j=1}^\infty g(\sigma e^{t^{1/3} a_j}) 
\right) \bigg] = \log {\rm Det}[ I - (1- e^{\beta \hat g_{t,\sigma}}) K_{\rm Ai}] 
\ee
$\hat g_{t,\sigma}(a)=g(\sigma e^{t^{1/3} a})$ (with $g(-e^x)=f(x)$ as defined in the introduction).
It leads to exactly the same expansion of $q_{t,\beta}(\sigma)$ in terms of the ${\cal L}_i$, i.e. all coefficients
being independent of the choice of the function $g(x)$,
where now 
\be \label{LLi} 
{\cal L}_i \equiv {\cal L}_i(\sigma) = \frac{\beta}{\pi} (\sigma \partial_\sigma)^{i+1} 
\int_0^{+\infty} dx \sqrt{x} g(\sigma e^{-x}) 
= \beta (\sigma \partial_\sigma)^{i} 
\int_{-\infty}^{+\infty} \frac{dp}{2 \pi} g(\sigma e^{-p^2}) 
\ee 
and the case of the KPZ equation is recovered for the choice $g(x)=g_{\rm KPZ}(x) = - \log(1-x)$. 
The important remark is thus that once the "initial condition"
\be \label{p0gen} 
p'_0(r) = {\cal L}_1(-e^{-r})
\ee 
is specified, then the hierarchy of equations for the $p_m(r)$ that we derived from the
simplified KP equation, yields exactly the same result for 
the $p_m(r)$ as a function of the ${\cal L}_i(\sigma=-e^{-r})$
as for the KPZ case. Hence it appears that for this more general
problem $\phi(t,r)= \partial_r^2 q_{t,\beta}(\sigma=-e^{-r})$ 
does satisfy also the KP equation, although with 
a different initial condition. This is shown here in Appendix \ref{app:fd} using
properties of FD.
Note that, strictly, this holds only if
the chosen function $g(x)$ is itself time-independent (hence, it does not apply a priori to the developments in Section V B in Ref. \cite{ShortTimeSystematic},
although we do find an application below, see also Appendix \ref{app:g}).

Of course, considering \eqref{remark1} is a natural extension, once
the conditions for a FD to lead to KP have been identified, i.e. some conditions
were given in Ref.
\cite{QRKP} (see also Appendix \ref{app:fd} here).
However the question of the initial condition was
not discussed there, and it is enlightening to see how it works out on the
small time expansion. 

\subsection{Large time large deviations}

\subsubsection{Leading order} 
\label{largelarge} 

We now study the large time limit, and search for a left tail large deviation form 
of the type \eqref{LDLT1}, in the limit where both $-r,t \to +\infty$ with $z=r/t <0$ fixed, that is
\be \label{LD0} 
\log \hat G(t,r) = \log  \langle \exp(- e^{ h(0,t) + \frac{t}{12} - r}) \rangle = - t^2 \Phi_-(z) + o(t^2)   \quad , \quad r = z t
\ee
where one must have $\Phi_-(z) \geq 0$ and $\Phi'_-(z) \leq 0$ since $\log \hat G$ is a negative increasing function of $r$.
Also one must have $\Phi_-(0)=0$ since the regime of typical fluctuations, $r \sim t^{1/3}$, correspond to $z=0$.
It is useful to note that the generating function can also be written as
\be
\log  \langle \exp(- e^{ H - r}) \rangle = {\rm Prob}(H + \gamma < r) 
\ee 
where $\gamma$ is a unit Gumbel variable, i.e. of CDF ${\rm Prob}(\gamma<a)=e^{-e^{-a}}$, independent
of $H$ (here $H=h(0,t) + \frac{t}{12}$). Hence in the limit $t \to + \infty$ one has $\log  \langle \exp(- e^{ H - z t}) \rangle \simeq \log {\rm Prob}(\frac{H}{t}<z)$ and \eqref{LD0} is equivalent to \eqref{LDLT1}.\\

Eq. \eqref{LD0} implies that we must search for the following scaling form for $\psi(t,r)$ in that limit 
\be
\psi(t,r) = \partial_r^2 \log \hat G(t,r) = H_0(\frac{r}{t}) \quad , \quad H_0(z) = - \Phi_-''(z) 
\ee
Substituting this form into the reduced equation \eqref{KP2}, we find that $H_0(z)$ 
must satisfy
\bea  \label{eqH0} 
\frac{1}{2} H_0(z) - z H_0'(z) + H_0(z) H_0'(z) = 0  
\eea 
where we can neglect the third derivative term $\frac{H_0'''(z)}{12 t^2}$ in the large time limit. 
One can also write the integrated version inserting $\hat \psi(r,t)=- t \hat \psi_0(H/t) +o(t)$ 
into \eqref{KP22}, with $\hat \psi_0(z)=- \Phi_-'(z)$ and $H_0(z)=\hat \psi_0'(z)$. The resulting
equation is $\frac{3}{2} \hat \psi_0 - z \hat \psi_0' + \frac{1}{2} (\hat \psi_0')^2=0$, i.e
the integrated version of \eqref{eqH0}, with the integration constant automatically fixed
from \eqref{KP22},
a constraint which can be written $- 3 \Phi'_-(0) + \Phi''_-(0)^2=0$. Given that $\Phi'_-(z)$
must be negative it implies that $\Phi'_-(0)=\Phi''_-(0)=0$. \\

The general solution of this equation with $H_0(0)=0$ and which is real for any $z<0$ reads
\be
H_0(z) = \frac{1}{A \pi^2} (1 - \sqrt{1 - A \pi^2 z}) = \frac{z}{2}+ \frac{A \pi^2}{8} z^2+O\left(z^3\right)
\ee
where $A>0$ is a constant, as yet undetermined. The factor $\pi^2$ has been introduced for convenience.
The $+$ branch is also solution but leads to $H(0)=2/(A \pi^2)$
and can be discarded. Note that $H_0(z)=z/2$ is also a solution, corresponding
to $A=0$. Let us now integrate twice to obtain $\Phi_-(z)$, using
that $\Phi_-(0)=\Phi'_-(0)=0$, leading to
\bea \label{phim} 
\Phi_-(z)= \frac{4}{15
   \pi ^6 A^3} \left( (1- A \pi ^2 z)^{5/2}-1\right) -\frac{z^2}{2 \pi ^2 A}+\frac{2 z}{3 \pi ^4 A^2} = - \frac{1}{12} z^3 + O(z^4)
\eea 
This is precisely the known result from the four methods \cite{sasorov2017large,JointLetter,SAOTsai2018,ShortTimeSystematic},
if we choose $A=1$. Note that the result at small $z<0$, $\Phi_-(z) \simeq_{z \to 0} - \frac{1}{12} z^3$, is obtained independently of the choice for $A$. This result can also be obtained \cite{LargeDevUs}
by matching with the cubic tail of
the Tracy Widom GUE distribution, from the regime of typical fluctuations
$r \sim t^{1/3}$, assuming no intermediate regime
(which seems to be indeed excluded by the present analysis). \\
%\be
% \Phi_-(z)= \frac{1}{30} \left(-8 A^3+20 A^2 z-15 A z^2+8 \sqrt{A}
%   (A-z)^{5/2}\right) \quad , \quad \Phi_-(z) \simeq - \frac{z^3}{12} + O(z^4) 
%\ee 
%At this stage the constant $A$ is still undetermined. If one compares
%with the known result 
%\bea
%\Phi_-(z)= -\frac{z^2}{2 \pi ^2 A}+\frac{2 z}{3 \pi ^4 A^2}+\frac{4
%   \left(\left(1- A \pi ^2 z\right)^{5/2}-1\right)}{15
%   \pi ^6 A^3}
%\eea 
%we see that the KP equation predicts it, provided we set $A=1/\pi^2$. 

Hence the KP equation provides a fifth and easy method to determine the large deviation function for
the left tail. The only remaining question is thus how to determine the constant $A$. It 
can be set by the outer tail $\Phi_-(z) \simeq \frac{4}{15 \pi} (-z)^{5/2}$ at large $z<0$
which can be obtained from the simplest and first term in the cumulant expansion
\cite{KrajLedou2018,JointLetter,ShortTimeSystematic,largedev}, which, by Jensen's inequality, is also an exact upper bound for
$\Phi_-(z)$. Equivalently,
the constant $A$ can be determined from the structure of the short time expansion (which is a
cumulant expansion). This was used in \cite{ShortTimeSystematic} to obtain the full function $\Phi_-(z)$. 
We will implement that program below
within the KP equation approach (in Section \ref{subsec:cumulants}), and 
show that indeed the constant $A=1$ by relating the short time and large time behaviors.

\subsubsection{Subleading orders} 
\label{sec:sub} 

We can now search for a systematic large time expansion in the large deviation regime, of the form
\be \label{rate} 
\log \hat G(t,r) = \log  \langle \exp(- e^{ h(0,t) + \frac{t}{12} - r}) \rangle = - t^2 \Phi_-(\frac{r}{t}) - t \, \Phi_1(\frac{r}{t}) - \Phi_2(\frac{r}{t}) + \dots
\ee
We thus insert in \eqref{KP2} the expansion
\be
\psi(t,r)= \partial_r^2 \log \hat G(t,r) = \sum_{m \geq 0} t^{-m} H_m(r/t)  \quad , \quad \hat \psi(t,r)= \partial_r \log \hat G(t,r)  = \sum_{m \geq 0} t^{1-m} \hat \psi_m(r/t) 
\ee
We find the following equation for $H_1(z)$
\bea
&& (H_0(z)-z) H_1'(z)+H_1(z) (H_0'(z)-\frac{1}{2}) = 0  
\eea 
and the integrated version $2 \hat \psi'_1 (z-\hat \psi_0)=\hat \psi_1$ (which fixes $\hat \psi_1(0)=- \Phi_1'(0)=0$). This 
leads to 
\be
H_1(z)=\frac{a_1}{\sqrt{1- A \pi^2 z}} \quad , \quad \hat \psi_1(z) = \frac{2 a_1}{A \pi^2} ( 1 - \sqrt{1- A \pi^2 z^2} ) 
\ee 
where $a_1$ is undetermined and $A$ was discussed in the previous Section. It leads to 
\be
\Phi_1(z) = \frac{4 a_1 }{3 \pi ^4
   A^2}(1- \left(1-\pi ^2 A z\right)^{3/2})-\frac{2 a_1 z}{\pi
   ^2 A} = - \frac{a_1}{2} z^2 + O(z^2) = -\frac{4 a_1 (-z)^{3/2}}{3 \pi  \sqrt{A}} + O(z) 
\ee 
where we have used that $\Phi_1(0)=0$. Indeed by matching to the typical regime 
we want $t \Phi_1(s t^{-2/3})$ to be at most of order unity at large $t$, hence 
$\Phi_1(z)$ is at most $|z|^{3/2}$ which implies $\Phi_1(0)=\Phi_1'(0)=0$.\\

However it turns out, as we see below from the examination of the cumulants in
the short time expansion, that in fact $a_1=0$ for the KPZ equation. It can already be guessed from
considering the subleading term in the small time expansion
\be
\log \hat  G(t,r) \simeq \frac{1}{\sqrt{4 \pi t}} {\rm Li}_{5/2}(- e^{-r}) = 
- \frac{1}{\sqrt{t}}  ( \frac{4}{15 \pi} (-r)^{5/2} + \frac{\pi}{6} \sqrt{-r} + \dots) 
\ee 
substituting $r= z t$, the first term is $O(t^2)$ and the second is $O(1)$, the
$O(1/t)$ term proportional to $(-z)^{3/2}$ is missing. We have used for $x \to +\infty$ and $\nu > 3/2$, see Ref. \cite{Math}
\be
{\rm Li}_{\nu}(- x) = - \frac{(\log x)^\nu}{\Gamma(\nu+1)} - \frac{\pi^2}{6} \frac{(\log x)^{\nu-2}}{\Gamma(\nu-1)} + \dots
\ee 
The complete argument for $a_1=0$ is made below. Note however that for the more
general problem defined in Remark 2 above, this constant is not necessarily zero.\\

Let us consider the next correction. Setting $a_1=0$, i.e. $H_1(z)=0$, the equation for $H_2(z)$ is
\be
H_2(z) (H_0'(z)-\frac{3}{2}) -z H_2'(z)+H_0(z) H_2'(z)+\frac{1}{12} H_0'''(z) = 0
\ee 
and its integrated version $- \frac{1}{2} \hat \psi_2 + ( \hat \psi'_0 - z)  \hat \psi'_2 + \frac{1}{12} \hat \psi_0'''=0$. 
We obtain the solution (from now on we set $A=1$, its value for the KPZ equation, see above Sections)
\be \label{resH2} 
H_2(z)
%=
%\frac{\pi ^4 \left(3 \sqrt{1-\pi ^2 z}-2\right)-96 a_2
%   \left(1-\pi ^2 z\right)^{3/2}}{96 \left(\pi ^2
%   z-1\right)^2 \left(\pi ^2 z+2 \sqrt{1-\pi ^2
%   z}-2\right)} 
   = \frac{96 a_2 \left(1-\pi ^2 z\right)^{3/2}+\pi ^4
   \left(2-3 \sqrt{1-\pi ^2 z}\right)}{96 \left(\pi ^2
   z-1\right)^2 \left(\sqrt{1-\pi ^2 z}-1\right)^2} \quad , \quad \hat \psi_2(z)=
     \frac{96 a_2 \left(\pi ^2 z-1\right)+\pi ^4}{48
   \pi ^2 \left(\pi ^2 z-1\right) \left(\sqrt{1-\pi ^2
   z}-1\right)} = \frac{\frac{1}{24}-\frac{4 a_2}{\pi
   ^4}}{z}+O\left(z^0\right)
   \ee
where $a_2$ is undetermined and we note that for general $a_2$ the function $\hat \psi_2(z) = - \Phi_2'(z)$
diverges at $z=0$. This leads to
\be
\Phi_2(z) = \frac{192 a_2 \sqrt{1-\pi ^2 z}-2 \left(\pi^4-96 a_2 \right) 
\log \left(\sqrt{1-\pi ^2 z} -1\right)+\pi ^4 \log \left(1-\pi ^2 z\right)}{48
   \pi ^4} + b 
\ee 
where $b$ is an (undetermined) integration constant. The constant $a_2$ is determined below from
the cumulant expansion and is found to be $a_2=\frac{\pi^4}{24}$. With this value of
$a_2$ we find that at small $z \to 0^-$ one has
\be \label{ll2} 
\Phi_2(z) \simeq \frac{1}{8}  \log (-z)+\frac{4+\log \frac{\pi ^6}{8}}{24}  + b -\frac{7
   \pi ^2 z}{96}+O\left(z^2\right)
\ee
One could try to match this result to the left tail asymptotics of the Tracy Widom GUE distribution,
recalled in \eqref{TWas} below, which gives, naively, $\log F_2(s=z t^{2/3}) \simeq - \frac{t^2}{12} |z|^3 
- \frac{1}{8} \log(|z| t^{2/3}) + C$. The first term is the correct leading small $|z|$ behavior of $\Phi_-(z)$, as discussed above, and the second term, i.e. the $- \frac{1}{8} \log |z|$ does agree indeed with \eqref{ll2}
and \eqref{rate}. Furthermore,
it is clear that $b$ could also be time dependent, and contain a $\log t$ term
(which disappears in taking the derivative $\hat \psi(t,r)$). This makes
the matching of the remaining $O(1)$ term delicate, i.e. the constant
part of $\Phi_2(z)$ at small $z$, especially in view of the following remark.
\\

{\bf Remark 3.} Having determined the subdominant rate functions for the generating
function in \eqref{rate}, we would like to translate it into large deviation rate functions
for the probability for $H$. Tentatively, one would search for a formula of the type, at large $t$,
\be \label{rate2} 
\log {\rm Prob}(H \leq z t)  \simeq - t^2 \Phi_-(z) - t \, \hat \Phi_1(z) - \hat \Phi_2(z) + \dots
\ee
While it is clear that the leading order involves the same function $\Phi_-(z)$ as in
\eqref{rate}, determining the subleading orders is more delicate. First, there is 
a useful upper bound from the Markov inequality
$\log {\rm Prob}(H \leq z t) = \log {\rm Prob}(e^{- e^{H- z t}} \geq \frac{1}{e})
\leq \log  \langle \exp(- e^{ H - z t}) \rangle + 1$. If the form \eqref{rate2} holds, it would imply
$\hat \Phi_1(z) \geq \Phi_1(z)=0$,
and $\hat \Phi_2(z) > \Phi_2(z)+1$. Second, if one follows
Ref. \cite{CorwinTails2} (Section 3.1) one has the lower bound
${\rm Prob}(H \leq z t) \geq \langle \exp(- e^{ H - \tilde z t}) \rangle
- e^{-e^{ t (z - \tilde z)}}$, for any $\tilde z$. The best one can do
is choose $\tilde z \simeq z - \frac{1}{t} \log (C t^2)$ and obtain a
bound of the type $\log {\rm Prob}(H \leq z t)  \geq - t^2 \Phi_-(\tilde z) - \hat \Phi_2(\tilde z) - a$
where $a>0$. Inserting $- t^2 \Phi_-(\tilde z)  \simeq - t^2 \Phi_-(z) + t \Phi_-'(z) \log (C t^2)$.
Hence it does not produce any useful lower bound on $\hat \Phi_1(z)$ (and even
suggests a possible $O(t \log t)$ term). \\

{\bf Remark 4.} In the very recent preprint \cite{CafassoTail} the following asymptotics is 
obtained by completely different methods
\be \label{cafasso1} 
\log \hat G(t,r) = - t^2 \Phi_-(\frac{r}{t}) - \frac{1}{6} \sqrt{1- \frac{\pi^2 r}{t}} + O(\log^2(|r|/t^{1/3})) + O(t^{1/3}) 
\ee
holds uniformly in $M^{-1} < t < M^{2/3} r$ as $s=- r/t^{1/3} \to +\infty$. In the region studied here,
both $-r,t \to +\infty$ with $r/t = z <0$ fixed, this agrees with our leading order $O(t^2)$ result (the function $\Phi_-(z)$ is the same). It also agrees with our conclusion that our $O(t)$ result, i.e. the function
$\Phi_1(z)$, vanishes. However it does not
allow for a comparison of our $O(1)$ result, since corrections terms in \eqref{cafasso1} are larger. 
However formula \eqref{cafasso1} also holds for $t$ fixed (e.g. of order $O(1)$) with
$r \to -\infty$, which goes beyond the results of this Section, but is discussed again
in Section \ref{sec:related}. 

\subsection{Matching small time and large time: the cumulants}
\label{subsec:cumulants} 

As we discovered above, inserting the expected scaling forms for the large time large deviation rate functions into the reduced KP equation, allows to determine their form up to a few unknown
constants ($A$, $c_1$, $c_2$ above). To determine those, we return to the small time
expansion, and pursue the program started in Ref. \cite{ShortTimeSystematic}. There the following expansion in cumulants was considered \cite{footnote1} 
(see Eqs. (129) to (132) there) 
\be
\log \hat G(t,r) = \sum_{n \geq 1} \frac{1}{n!} \, \kappa_n(t,r)
\ee 
where $\log \hat G(t,r)=q_{t \beta}(\sigma=-e^{-r})$ there, for the choice $g(x)=g_{\rm KPZ}(x)$, $\beta=1$,
but it holds more generally, see Remark 2. The cumulant $\kappa_n(t,r)$ re-groups all terms of degree $n$ in 
the expansion studied above, i.e. it is by definition a homogeneous polynomial
of total degree $n$ in $p'_0(r) = {\cal L}_1$ and its derivatives $- \partial_r {\cal L}_i={\cal L}_{i+1}$. 
Each cumulant has the following short time series expansion in $t$ 
\be
\kappa_n(t,r) = t^{\frac{n}{2}-1} \kappa_n^0(r) + \sum_{p \geq 1} t^{\frac{n}{2}-1+p} \kappa_n^p(r)
\ee
The form of the leading term was found exactly in Ref. \cite{ShortTimeSystematic} (see below) and shown to sum up to produce the large time large deviation function $\Phi_-(z)$ studied above. Here we recover the result
for $\kappa_n^0(r)$ from the KP equation, and obtain the next subleading term $\kappa_n^1(r)$.

Let us recall the iterative solution for the functions $p'_m(r)$ defined in \eqref{serpsi}
corresponding to the term of order $t^{\frac{m-1}{2}}$ in the small time expansion 
of $\psi(t,r) = \partial_r^2 \log \hat G(t,r)$. We see
from e.g. \eqref{hier} (see also Appendix \ref{app:shorttime}) 
that the structure of the result for the $p_m$ is a polynomial in $p'_0$ and derivatives of the form
(schematically, we only indicate the degree in $p'_0$ and the total number of derivatives)
\be \label{scheme} 
p'_m = (\partial_r)^{m} (p'_0)^{m+1} + (\partial_r)^{m+1}  (p'_0)^{m-1} + \dots 
\ee
The leading term corresponds to $\kappa^0_{m+1}$, the second one to $\kappa^1_{m-1}$, and so on.
Below we will denote $p'_m = p^{0 \prime}_m + p^{1 \prime}_m + \dots$ the above decomposition.\\

We will thus introduce the following expansion for $\psi(t,r)$
\be
\psi(t,r) = \psi_0(t,r) + \psi_1(t,r) + \dots \quad , \quad \psi_p(t,r)= \partial_r^2 \sum_{n \geq 1} \frac{\kappa_n^p(t,r)}{n!}
\ee
where $\psi_0(t,r)$ corresponds to $\kappa_n^0$, $\psi_1(t,r)$ corresponds to $\kappa_n^1$, and so
on. This expansion corresponds to treating perturbatively the third derivative term in the reduced KP equation \eqref{KP2} and can be called {\it semi-classical expansion}. The leading term is thus obtained by neglecting the third derivative term in \eqref{KP2}, i.e. solving 
\be \label{KP2n2} 
 \partial_t \psi_0 +  \psi_0 \partial_r \psi_0 +  \frac{1}{2 t}  \psi_0 =0
\ee 
This is precisely the equation which was solved in Section \ref{largelarge}, but only 
in the large $t$, large $r<0$ limit, using the large deviation ansatz $\psi(t,r) \simeq H_0(r/t)$ with $H_0(z)=- \Phi_-''(z)$
leading to Eq. \eqref{eqH0}. Here we provide a more complete solution, 
valid at all time $t$ and $r$. We will do it by two methods (i) a series expansion,
allowing to recover the result of \cite{ShortTimeSystematic} (ii) a mapping to
Burgers equation, which is reminiscent of what was found in \cite{largedev}
and allows to calculate easily the next subleading term. Our result leads to a better understanding
of how the small time and large time large deviations are related. 

The first method is to search for a solution of \eqref{KP2n2} as a series expansion
at small time of the form $\psi_0(r,t)=\sum_{m \geq 0} p^{0\prime}_m(r) t^{\frac{m-1}{2}}$. This leads to
the simplified recursion for $m \geq 0$
\bea
\frac{m+1}{2} p^0_{m+1} + \frac{1}{2} \sum_{m_1+m_2=m, m_1 \geq 0, m_2 \geq 0} p^{0\prime}_{m_1} p^{0\prime}_{m_2} = 0
\eea 
and one obtains exactly the $p^{0\prime}_m$, $m \geq 0$ as a function of $p'_0$ as
\bea \label{res11} 
p^{0\prime}_m(r) = \frac{2^m}{(m+1)!} (-1)^m (\partial_r)^{m} (p'_0(r))^{m+1}
\eea 
using the identity for $m \geq 0$ for any function $f(r)$
\be
\partial_r^m f^{m+2} = \frac{1}{2 (m+1)}  \sum_{m_1+m_2=m, m_1 \geq 0 , m_2 \geq 0} 
\frac{(m+2)!}{(m_1+1)! (m_2+1)!} \partial_r^{m_1} f^{m_1+1} \partial_r^{m_2} f^{m_2+1}
\ee 
which is easily checked with Mathematica. The result \eqref{res11} is exactly equivalent to
the formula (132) in Ref. \cite{ShortTimeSystematic}, more precisely $p^{0\prime}_m(r)$ equals
$\partial_r^2 \frac{\kappa_n^0(r)}{n!}$ with $n=m+1$, using $\sigma=-e^{-r}$, $\sigma \partial_\sigma=-\partial_r$ and
${\cal L}_1=p_0'$. Obtaining the subleading term with that method is a bit tedious, so we
now consider an equivalent, but more convenient method.\\

Let us first note that if we perform the following change of variable
\be
\psi(t,r) = \frac{1}{\sqrt{t}} \tilde \psi(\sqrt{t},r) 
\ee 
then the reduced KP equation \eqref{KP2} becomes, for the function $\tilde \psi(u,r)$
\be \label{KPw} 
\partial_u \tilde \psi(u,r) + \partial_r (\tilde \psi(u,r)^2) + \frac{1}{6} u \partial_r^3 \tilde \psi(u,r) = 0 \quad , \quad u=\sqrt{t} 
\ee 

We can now again perform the perturbative expansion in the cubic derivative term and write, for $p=0,1,..$ 
\be \label{sum0} 
\psi(t,r) = \sum_{p \geq 0} \psi_p(t,r) \quad , \quad \psi_p(t,r) = \frac{1}{\sqrt{t}} \tilde \psi_p(\sqrt{t},r) 
\ee 
The leading term then obeys the Burgers equation
\be
\partial_u \tilde \psi_0(u,r) + \partial_r (\tilde \psi_0(u,r)^2)  = 0 
\ee 
The general solution can be expressed as $F(\tilde \psi_0(u,r) , r - 2 u \tilde \psi_0(u,r))=0$ for some
function $F(x,y)$.
The solution of interest here has "initial condition" $\tilde \psi_0(u=0,r)=p'_0(r)$. It is
thus given by
\be \label{solub} 
\tilde \psi_0(u,r) = p_0'(r- 2 u \tilde \psi_0(u,r)) \Rightarrow \psi_0(t,r) = \frac{1}{\sqrt{t}} p_0'(r - 2 t \psi_0(t,r)) 
\ee
By expanding at small $t$ one can check that it produces the leading order of the cumulants as given above
in \eqref{res11}. \\

The result \eqref{solub} is quite general, see Remark 2. In the present case of the KPZ equation, using \eqref{init}, 
$\psi_0(t,r)$ is solution of
\be \label{sc} 
\psi_0(t,r) = \frac{1}{\sqrt{4 \pi t}} {\rm Li}_{1/2}(- e^{- r + 2 t \psi_0(t,r)}) 
\ee 
where we recall $\psi_0(t,r) = \partial_r^2 \sum_n t^{\frac{n}{2}-1} \frac{\kappa_n^0(r)}{n!}$ is the summation of the leading term in the small time expansion of each cumulant. 

As was noticed in \cite{ShortTimeSystematic} for large negative $r \sim z t$, $z<0$, the leading term in each cumulant
$\kappa_n^0(r)$ allows to obtain the large deviation function $\Phi_-(z)$. From the large negative $r$ asymptotics
of the polylogarithm \cite{Math} 
one has
\be \label{expLi} 
p'_0(r)=  \frac{1}{\sqrt{4 \pi}} {\rm Li}_{1/2}(-e^{-r}) = - \frac{1}{\pi} (-r)^{1/2} + \frac{\pi}{24} (-r)^{-3/2} + \dots 
\ee
Inserting into \eqref{sc} we obtain that the following limit exists, for $z<0$
\be
H_0(z) = \lim_{t \to +\infty} \psi_0(t,z t) 
\ee
and $H_0(z)$ obeys the self-consistent equation
\be
H_0(z) = - \frac{1}{\pi} \sqrt{-z + 2 H_0(z)} \quad  \Rightarrow \quad H_0(z) =\frac{1}{\pi^2} (1 \pm \sqrt{1- \pi^2 z}) 
\ee 
The correct branch (actually reached in the large time limit) is $-$ and one recovers 
the result of the previous Section for the large deviation function $H_0(z)$, with $A=1$. This method thus relates the
small and large time and allows to determine the missing constant $A=1$ for the large time
large deviation rate function.\\

Let us now briefly consider the general case of the linear statistics discussed in Remark 2.
For a general function $g(x)$ the leading term in the semi-classical expansion is 
given by \eqref{solub}, where the function $p'_0(r)$ is given by \eqref{p0gen}
and \eqref{LLi}. This case is studied in detail in Appendix \ref{app:g},
where a connection with the Section V B in Ref. \cite{ShortTimeSystematic}
and with the self-consistent equation in Ref. \cite{largedev} is obtained.
It shows that there are other large time large deviation solutions of the KP 
equation, of the form $\psi(t,r) \simeq t^{\alpha-1} H_0(r/t^\alpha)$, with
continuously varying exponent $\alpha$ and function $H_0$, which corresponds
to other functions $g(x)$ in Remark 2 (and the "monomial walls" in the Coulomb gas terminology
of Ref. \cite{largedev}). The case of the KPZ equation 
is recovered for $\alpha=1$.\\

Let us now go back to the KPZ equation and study the subleading term. Inserting \eqref{sum0} into \eqref{KPw} we obtain the
following equation for $\tilde \psi_1(u,r)$
\be
\partial_u \tilde \psi_1(u,r) + 2 \partial_r [\tilde \psi_1(u,r) \tilde \psi_0(u,r)] + \frac{1}{6} u \partial_r^3 \tilde \psi_0(u,r) = 0
\ee
It is more convenient to write $\tilde \psi_1(u,r) = \partial_r \phi(u,r)$, integrate once w.r.t. $r$ and obtain the equation for $\phi$
\be \label{eqphi} 
\partial_u \phi(u,r) + 2 \tilde \psi_0(u,r) \partial_r \phi(u,r) = - \frac{1}{6} u \partial_r^2 \tilde \psi_0(u,r) 
\ee 
which can be seen as a convection equation for a passive scalar $\phi(u,r)$ in the Burgers velocity field
$2 \tilde \psi_0(u,r)$. It is easy, and useful for later checks, to first extract the small $u$ expansion of $\phi(u,r)$ from
\eqref{eqphi} and \eqref{solub}. One finds
\bea \label{expu} 
&& \phi(u,r) = - \frac{1}{12} u^2 \partial_r^2 p_0'(r)+ u^3 \partial_r [ \frac{1}{6} p_0{}^{(3)}(r) p_0'(r)+\frac{1}{12}
   p_0''(r){}^2 ] \\
   && + u^4 \partial_r [ -\frac{2}{3} p_0{}^{(3)}(r) p_0'(r)
   p_0''(r)-\frac{1}{6} p_0{}^{(4)}(r)
   p_0'(r){}^2-\frac{1}{9} p_0''(r){}^3 ] + O(u^5) 
\eea
To solve the equation \eqref{eqphi} it is then natural to work in the variable $u,\tilde \psi_0$ instead of $u,r$. Indeed one has, from 
the solution \eqref{solub} to Burgers equation
\be \label{ru0} 
r = r(u,\tilde \psi_0) = 2 u \tilde \psi_0 + g(\tilde \psi_0) \quad , \quad p_0'(g(a))=a 
\ee 
where $g$ is the inverse function of $p_0'$. Let us define
\bea
\tilde \phi(u,\psi_0) = \phi(u,r(u,\tilde \psi_0) = \phi(u, 2 u \tilde \psi_0 + g(\tilde \psi_0)) 
\eea 
We now obtain an equation for this function.
Taking a derivative w.r.t. $u$, one obtains, using the equation \eqref{eqphi} for $\phi$
\bea \label{ww} 
\partial_u \tilde \phi(u,\tilde \psi_0) = \partial_u \phi(u, 2 u \tilde \psi_0 + g(\tilde \psi_0)) + 2 \tilde \psi_0 \, 
\partial_r \phi(u, 2 u \tilde \psi_0 + g(\tilde \psi_0)) = - \frac{1}{6} u \partial_r^2 \tilde \psi_0(u,r)
\eea 
Let us evaluate the r.h.s. One has from \eqref{ru0}, by variation
\bea \label{rdiff} 
\partial_r \tilde \psi_0 = \frac{1}{\partial_{\tilde \psi_0} r} = \frac{1}{2 u + g'(\tilde \psi_0)} 
\eea 
Taking a derivative $\partial_r$ and using again \eqref{rdiff}, one finally obtains from \eqref{ww}
\bea
\partial_u \tilde \phi(u,\tilde \psi_0) = \frac{1}{6} u \frac{g''(\tilde \psi_0) \partial_r \tilde \psi_0}{(2 u + g'(\tilde \psi_0))^2} 
= \frac{1}{6} \frac{u g''(\tilde \psi_0)}{(2 u + g'(\tilde \psi_0))^3}
\eea 
One can then easily integrate this equation w.r.t. $u$, with the constraint that $\phi$ must vanish at $u=0$
from \eqref{expu}. One obtains
\bea
\tilde \phi(u,\tilde \psi_0) = \frac{g''(\tilde \psi_0) u^2}{12 g'(\tilde \psi_0) (2 u + g'(\tilde \psi_0))^2} 
%= - \frac{1}{12} u^2 p_0'''(g(\psi_0))  + \dots = - \frac{1}{12} u^2 p_0'''(g(p'_0))  + \dots 
%= - \frac{1}{12} u^2 p_0'''(r)  + \dots 
\eea 
From the definition of the inverse function $g(a)$ one can now use the relations
\be
g'(a) = \frac{1}{p_0''(g(a))} \quad , \quad g''(a) = - g'(a)^2 \frac{p_0'''(g(a))}{p_0''(g(a))}
\ee 
and we obtain our final result for the two first orders, summarized as follows
\bea \label{fullsolu} 
&& \psi(t,r) = \frac{1}{\sqrt{t}} [ \tilde \psi_0(u,r) + \partial_r \phi(u,r) + \dots] \quad , \quad u=\sqrt{t} \quad , \quad 
\tilde \psi_0(u,r) = p_0'(r- 2 u \tilde \psi_0(u,r))
\\
&& \phi(u,r) = - \frac{u^2}{12} \frac{p_0'''(r - 2 u \tilde \psi_0(u,r))}{(1 + 2 u p_0''(r - 2 u \tilde \psi_0(u,r)) )^2} 
\quad , \quad
\tilde \psi_0(u,r) = \sum_{ m \geq 0} u^m \frac{2^m}{(m+1)!} (-1)^m (\partial_r)^{m} (p'_0(r))^{m+1}
\eea 
It is easy to expand this result in powers of $u=\sqrt{t}$ and recover the
result \eqref{expu}, which provides a check on our exact solution.\\

We now consider again the limit of large $t$, large negative $r$, with $z=r/t <0$ fixed.
Up to higher order terms, $O(1/t^3)$, we only need the (semi-classical) expansion in the cubic derivative
of the KP equation as
$\psi(t,r) \simeq \psi_0(t,r) + \psi_1(t,r)$.
We will now find that $\psi_0(t,r) \simeq H_0(z) + \frac{1}{t^2} H_2^{(2)}(z)$
and $\psi_1(t,r) \simeq \frac{1}{t^2} H_2^{(1)}(z)$. Hence (i) there are no
correction of order $O(1/t)$, i.e. the function $H_1(z)$ is zero, as claimed
in the previous Section (ii) two pieces add up to give the total subleading rate
function $H_2(z) = H_2^{(1)}(z) + H_2^{(2)}(z)$.\\

To evaluate $\psi_1$ we use the large negative $r$ asymptotics. 
\bea
p'_0(r) = - \frac{1}{\pi} (-r)^{1/2} \quad , \quad p''_0(r) =  \frac{1}{2 \pi} (-r)^{-1/2}
\quad , \quad p'''_0(r) =  \frac{1}{4 \pi} (-r)^{-3/2}
\eea 
Inserting in the solution \eqref{fullsolu} we obtain
\bea
 \psi_1(u,r) &=& - \frac{\sqrt{t}}{12} \frac{1}{4 \pi} 
\partial_r \bigg[ (- r + 2 t H_0)^{-3/2} 
\frac{1}{(1 + \frac{1}{\pi} \sqrt{t} (-r - 2 t H_0)^{-1/2} )^2} \bigg] \\
%&& = - \frac{1}{12 t^2} \frac{1}{4 \pi}  
%\partial_z \bigg[ (- z + 2 H_0)^{-3/2} 
%\frac{1}{(1 + \frac{1}{\pi} (-z - 2 H_0)^{-1/2} )^2} \bigg]
%\eea
%We can now use that $- \pi H_0= \sqrt{-z + 2 H_0}$ and obtain
%\bea
%&& \psi_1(u,r) = \frac{1}{12 t^2} \frac{1}{4 \pi}  
%\partial_z \bigg[ \frac{1}{\pi^3 H_0^3} 
%\frac{1}{(1 - \frac{1}{\pi^2 H_0}  )^2} \bigg]
&=& \frac{1}{t^2} H^{(1)}_2(z) 
%\partial_z \bigg[ \frac{1}{ H_0} 
%\frac{1}{(\pi^2 H_0 - 1  )^2} \bigg]  
\quad , \quad H^{(1)}_2(z) = \frac{1}{48} 
\partial_z \bigg[ \frac{1}{ H_0(z)} 
\frac{1}{(\pi^2 H_0(z) - 1  )^2} \bigg] 
\eea
where we have used that $- \pi H_0(z)= \sqrt{-z + 2 H_0(z)}$ and 
we recall that $H_0(z)=\frac{1}{\pi^2} (1 - \sqrt{1- \pi^2 z})$.\\

Inserting now the large $r$ expansion \eqref{expLi}, and keeping the subdominant term
in the equation for $\psi_0$, i.e. $\psi_0 = \frac{1}{\sqrt{t}} p_0'(r - 2 t \psi_0)$,
we obtain $\psi_0=H_0 + \frac{1}{t^2} H^{(2)}_2$ with 
%
%\bea
%&& \psi = \frac{1}{\sqrt{t}} p_0'(r - 2 t \psi) \\
%&& p'_0(r)=  \frac{1}{\sqrt{4 \pi}} {\rm Li}_{1/2}(-e^{-r}) = - \frac{1}{\pi} (-r)^{1/2} + \frac{\pi}{24} (-r)^{-3/2} + \dots 
%\eea 
%Solving with $\psi=H_0 + \frac{1}{t^2} H_2$ we obtain
\be
H^{(2)}_2(z) = \frac{1}{24 H_0(z)^2 (1- \pi^2 H_0(z))} 
\ee 
Adding the two contributions we find
\be \label{resH2n} 
H_2(z)= \frac{1}{24 H_0(z)^2 (1- \pi^2 H_0(z))}  + \frac{1}{48} 
\partial_z \bigg[ \frac{1}{ H_0(z)} 
\frac{1}{(\pi^2 H_0(z) - 1  )^2} \bigg] \quad , \quad H_0(z)=\frac{1}{\pi^2} (1 - \sqrt{1- \pi^2 z})
\ee
One can now check that this result is identical to the one obtained in
\eqref{resH2} from the large time large deviation ansatz, provided one
sets $a_2=\frac{\pi^4}{24}$. In fact the second term in \eqref{resH2n} corresponds to 
setting $a_2=0$, while the first is the one
proportional to $a_2$ (i.e. the solution of the homogeneous part of the equation for $H_2$). 
Once again, the calculation of the cumulants from the short-time expansion 
allows to fix the unknown constant $a_2$ in the large time large deviation 
subleading rate function.

\subsection{A related expansion: intermediate times}
\label{sec:related}

One can consider a related expansion which allows a systematic study of the
tail of $\log \hat G(t,r)$ and of its derivative $\hat \psi(t,r)=\partial_r \log \hat G(t,r)$ (which is solution of the
integrated version \eqref{KP22} of the reduced KP equation) at large $r<0$ for any time $t$.

Consider again the small time series expansion \eqref{serpsi}.
We now assume that the functions $p_m(r)$ have the following 
expansion $p_m(r) = \sum_{n \geq 0} p_{m,n}  (-r)^{\frac{3-n}{2}}$
for $r \to -\infty$. This is certainly the case for the KPZ equation,
and for some class of FD as in Remark 2 (the more general 
class studied in Appendix \ref{app:g} can be studied 
by similar series expansions involving different exponents). Hence we look for a solution as a double series
\be \label{ser22} 
\hat \psi(t,r)= \sum_{0 \leq m \leq n} p_{m,n} t^{\frac{m-1}{2}} (-r)^{\frac{3-n}{2}} = 
\sum_{m \geq 0} p_{m}(r) t^{\frac{m-1}{2}} = 
\sum_{n \geq 0} q_{n}(t) (-r)^{\frac{3-n}{2}}
\ee 
since, as we find below, $p_{m,n}=0$ for $m > n$. The coefficients $p_{m,n}$ encode information for several limits.
(i) First, the leading small time behavior for $t \to 0$ at fixed $r<0$ (for $m=0$) is given by
\be \label{st0} 
\hat \psi(t,r) \simeq p_0(r) =  \frac{1}{\sqrt{t}} \sum_{n \geq 0} p_{0,n} (-r)^{\frac{3-n}{2}} = - \frac{1}{\sqrt{4 \pi}} {\rm Li}_{3/2} (- e^{-r}) 
\ee 
where the last equality is valid only for the particular case of the KPZ equation, see \eqref{init}.
(ii) Second, the leading large time large deviations \eqref{LD0} for $r=z t <0$ and $t \to +\infty$ (for $m=n$) is given by
\be \label{eqeq} 
\hat \psi(t,r) \simeq t \sum_{n \geq 0} p_{n,n} (- \frac{r}{t})^{\frac{3-n}{2}} = - t \, \Phi'_-(z=\frac{r}{t})
\ee 
(iii) Finally, the series \eqref{ser22} contain the information about the
large $r<0$ expansion at fixed time $t$, encoded in the functions $q_n(t)$, i.e.
\be \label{fixedt} 
\hat \psi(t,r)=\sum_{n \geq 0} q_{n}(t) (-r)^{\frac{3-n}{2}} \quad , \quad q_n(t) = \sum_{m=0}^{n} p_{m,n} 
t^{\frac{m-1}{2}}
\ee
We will determine below some of these functions $q_n(t)$. To this aim one can note
that they satisfy differential equations which can be solved recursively. We find it
easier however to study instead the recursion for the $p_{m,n}$. 
Inserting in \eqref{KP22} we obtain recursion relations which show that 
all $p_{m >n,n}=0$. The coefficients
$p_{0,n}$ for $n \geq 0$ are arbitrary (i.e. determined as in the previous Sections by the function $p_0(r)$).
All $p_{m,n}$ with $m \geq 1$ and general $n \geq 1$
can be obtained from the set of $p_{0,n}$ as follows
\be
p_{m,n} = \frac{-2}{m} \big( \frac{(n-9) (n-7) (n-5)}{96}  p_{m-2,n-6} + \sum_{n_1=0}^{n-1} 
\frac{(3-n_1)(4+n_1-n)}{8} \sum_{m_1=\max(0,m-n+n_1)}^{\min(n_1,m-1)} p_{m_1,n_1} 
p_{m-1-m_1,n-1-n_1} \big)
\ee 
%\be
%\frac{m}{2} p_{m,n} +\frac{(n-9) (n-7) (n-5)}{96}  p_{m-2,n-6} 
%+ \sum_{m_1,n_1,m_2,n_2 \geq 0} \frac{(3-n_1)(3-n_2)}{8} p_{m_1,n_1} p_{m_2,n_2} 
%\delta_{m_1+m_2,m-1} \delta_{n_1+n_2,n-1} =0
%\ee 
where all $p_{m,n}=0$ for either $m < 0$ or $n < 0$. 
This recursion is easily generated using Mathematica. One finds that 
$p_{1,1}= - \frac{9}{4} p_{0,0}^2$ and so on. 
If we suppress the first term on the r.h.s, which arises from the cubic derivative in the KP equation,
one can check that one reproduces the expansion given in \eqref{res11}.\\
%\bea \label{res112} 
%p^{0\prime}_m(r) = \frac{2^m}{(m+1)!} (-1)^m (\partial_r)^{m} (p'_0(r))^{m+1}
%\eea 
%

We now specialize to the KPZ equation, for which the $p_{0,n}$ are 
determined from \eqref{st0}. Let us recall the expansion, for $r \to -\infty$ 
(here we need only the formula for $s$ a positive half-integer)
\be
{\rm Li}_s(- e^{-r}) = -2 \sum_{k=0}^{+\infty} (1-2^{1-2 k}) \zeta(2k)  \frac{(-r)^{s-2 k} }{\Gamma(s+1-2 k)} 
\ee 
Hence $p_{0,n}=0$ unless $n$ is a multiple of $4$ and
\be
p_{n=4 k} = \frac{1}{\sqrt{\pi}} \frac{(1-2^{1-\frac{n}{2}}) \zeta(\frac{n}{2})}{\Gamma(\frac{5-n}{2})} 
\ee 
The above recursion then leads to the following solutions for the lowest $q_n(t)$ up to $n=12$
\bea
 && q_0(t) = \frac{2}{3 \pi  \sqrt{t}} 
 \quad , \quad 
 q_1(t) = -\frac{1}{\pi ^2} \quad , \quad 
 q_2(t) = \frac{\sqrt{t}}{\pi ^3} \quad , \quad 
 q_3(t) = -\frac{2 t}{3 \pi ^4} \quad , \quad
  q_4(t) = \frac{t^{3/2}}{4 \pi ^5}+\frac{\pi }{12
   \sqrt{t}} \\
   &&
 q_5(t) = \frac{1}{12} \quad , \quad
 q_6(t) = \frac{\sqrt{t}}{48 \pi }-\frac{t^{5/2}}{24 \pi ^7} 
 \quad , \quad q_7(t) = -\frac{t}{48 \pi ^2} 
\quad , \quad
q_8(t) = \frac{t^{7/2}}{64 \pi ^9}+\frac{7 \pi ^3}{960 \sqrt{t}} 
\\
&&  q_9(t) = \frac{t^2}{48 \pi ^4}+\frac{5 \pi ^2}{144} \quad , \quad 
 q_{10}(t) = -\frac{t^{9/2}}{128 \pi ^{11}}-\frac{t^{5/2}}{384 \pi ^5}+\frac{5 \pi 
   \sqrt{t}}{72} \quad , \quad
 q_{11}(t) = \frac{13 t}{192}-\frac{t^3}{48 \pi ^6} \\
&&  q_{12}(t) = \frac{7 t^{11/2}}{1536 \pi ^{13}}+\frac{5 t^{7/2}}{1536 \pi ^7}+\frac{259
   t^{3/2}}{9216 \pi }+\frac{31 \pi ^5}{4608 \sqrt{t}} 
\eea 
One can verify that, keeping only the leading term at large $t$ for 
each $q_n(t)$, i.e. $p_{n,n} t^{\frac{n-1}{2}}$, agrees, as it should according to \eqref{eqeq}, 
with the expansion of $- t \Phi_-'(z)$ in $z=r/t$ at large $z<0$, 
from the solution \eqref{phim} (with the correct value $A=1$), a rather non-trivial check
\bea
t \hat \psi_0(r/t)= - t \, \Phi_-'(\frac{r}{t}) &=&
\frac{2 (-r)^{3/2}}{3 \pi  \sqrt{t}}-\frac{- r}{\pi
   ^2}+\frac{\sqrt{-r} \sqrt{t}}{\pi ^3}-\frac{2 t}{3
   \pi ^4}+\frac{\sqrt{-\frac{1}{r}} t^{3/2}}{4 \pi
   ^5}-\frac{\left(-\frac{1}{r}\right)^{3/2}
   t^{5/2}}{24 \pi
   ^7} \\
   & +& \frac{\left(-\frac{1}{r}\right)^{5/2}
   t^{7/2}}{64 \pi
   ^9}-\frac{\left(-\frac{1}{r}\right)^{7/2}
   t^{9/2}}{128 \pi ^{11}}+\frac{7
   \left(-\frac{1}{r}\right)^{9/2} t^{11/2}}{1536 \pi
   ^{13}}+O((-\frac{1}{r})^{11/2})
\eea
Furthermore, since $p_{n-2,n}=0$ for all $n \geq 2$, we immediately see that the subleading function $H_1(z)=\hat \psi_1'(z)$
studied in Section \ref{sec:sub} is indeed zero, and that the above results are 
consistent with the series expansion of the next subleading function  
$H_2(z)=\hat \psi_2'(z)$ obtained in that Section, i.e. one can check that 
\be
\hat \psi_2(z) = \sum_{n \geq 4} p_{n-4,n} (-z)^{\frac{3-n}{2}} 
\ee 
is indeed the function for $z<0$ found in \eqref{resH2} with the correct value $a_2=\frac{\pi^4}{24}$.\\

We can now integrate \eqref{fixedt} over $r$ to obtain the series expansion 
for large $r<0$
\bea \label{finfin} 
&& \log \hat G(t,r) 
%= - \sum_{0 \leq m \leq n} \frac{2 p_{m,n} }{5-n}  t^{\frac{m-1}{2}} (-r)^{\frac{3-n}{2}} 
=
- \sum_{n \geq 0, n \neq 5} q_n(t) \frac{2}{5-n} (-r)^{\frac{5-n}{2}} 
- \frac{1}{12} \log( -r) + Q(t) \\
&& = 
-\frac{4 (-r)^{5/2}}{15 \left(\pi 
   \sqrt{t}\right)}+\frac{r^2}{2 \pi ^2}-\frac{2
   (-r)^{3/2} \sqrt{t}}{3 \pi ^3}-\frac{2 r t}{3
   \pi ^4}-\frac{\sqrt{-r} \left(3 t^2+\pi ^6\right)}{6
   \left(\pi ^5 \sqrt{t}\right)}- \frac{1}{12} \log(-r) + Q(t) 
- \sum_{n \geq 6} q_n(t) \frac{2}{5-n} (-r)^{\frac{5-n}{2}} \nn
\eea
Here $Q(t)$ is an undetermined integration constant of $O(1)$ in the large $r<0$ expansion.
Note that all terms with a positive power of $r$ appearing in \eqref{finfin} are already contained
in the function $-t^2 \Phi_-(r/t)$, apart from the term
$-  \frac{\pi}{6} \sqrt{-r/t}$. This additional term is consistent with the one discussed in Remark 4. \\

{\bf Remark 5.} For the KPZ equation many of the $p_{m,n}$ vanish. Indeed they vanish if $m-n$ is not
a multiple of $4$. The series 
has the following structure
\be \label{ser223} 
\hat \psi(t,r)= \sum_{0 \leq q \leq k} [ p_{4 q ,4 k} t^{-\frac{1}{2} + 2 q} (-r)^{\frac{3}{2}-2 k} 
+ p_{4 q +1 ,4 k +1 } t^{2 q} (-r)^{1-2 k} + p_{4 q+2 ,4 k+2} t^{\frac{1}{2} + 2 q} (-r)^{\frac{1}{2}-2 k} 
+ p_{4 q+3 ,4 k+3} t^{1 + 2 q} (-r)^{-2 k} ]
\ee 
Hence it naturally splits in the sum of four functions, for which one can also
obtain coupled series recursion relations. 
 
\subsection{Large time expansion, typical fluctuations}

For completeness we now address briefly the regime of typical fluctuations
in the large time limit. Not surprisingly, once the scaling form is introduced, it reproduces 
the KPZ fixed point result of Ref. \cite{QRKP}. However it allows in principle to 
study the finite time corrections. \\

In the large time and typical fluctations regime, we expect that the generating function
\eqref{defpsi}  takes the form
\be \label{typtyp} 
\hat G(t,r) = P_0(r/t^{1/3}) + t^{-a} P_1(r/t^{1/3}) + \cdots  \quad , \quad P_0(s)= \lim_{t \to +\infty} {\rm Prob}(\frac{h(0,t)+\frac{t}{12}}{t^{1/3}} <s)
\ee 
where the last equality follows by construction of the generating function. At this stage we allow some
freedom for the decay exponent $a$ of the subleading corrections (see below). 
We thus look for a solution to the reduced KP equation \eqref{KP2} of the form 
\be
\psi(t,r) =  \partial_r^2 \log \hat G(t,r) = t^{-2/3} ( \psi_0(r/t^{1/3}) + t^{-a} \psi_1(r/t^{1/3}) + \cdots )
\ee 
The function $\psi_0(s)$ must satisfy
\be \label{psi0} 
12 \psi _0 \psi _0' -4 s \psi _0' -2 \psi_0 +\psi''' _0 = 0 
\ee
Note that one can also consider the integrated version which, from \eqref{KP22} leads to
\bea \label{hatpsi1} 
&& \psi(t,r) =  \partial_r \log \hat G(t,r) = t^{-1/3} ( \hat \psi_0(r/t^{1/3}) + t^{-a} \hat \psi_1(r/t^{1/3}) + \cdots ) \\
&& 6 (\hat \psi'_0)^2 - 4 s \hat \psi'_0 + 2 \hat \psi_0 + \hat \psi_0'''=0
\eea
with $\psi_0 = \hat \psi_0'$. \\

A solution to the equation \eqref{psi0} is obtained from a solution $q(s)$ to the Painleve II equation as
\be \label{qq} 
\psi_0(s) = - q(s)^2 \quad , \quad q'' = s q + 2 q^3 
\ee 
This is verified by inserting into \eqref{psi0} leading to
\be
12 \psi _0 \psi _0' -4 s \psi _0' -2 \psi_0 +\psi''' _0  |_{\psi_0=-q^2} = -  (2 q \frac{d}{ds} + 6 q') (q'' - s q - 2 q^3) = 0
\ee
This solution, together with $q(s) \sim - {\rm Ai}(s)$ for $s \to +\infty$, corresponds to the TW-GUE 
distribution $F_2(s)$ 
\be
\partial_s^2 \log P_0(s) = - q(s)^2  \quad , \quad P_0(s) = e^{- \int_s^{+\infty} du (u-s) q(u)^2} = F_2(s)
\ee 
This is the standard analysis, also obtained from the KP equation satisfied by the KPZ fixed point in Ref. \cite{QRKP}.\\

Let us recall the large negative $s$ asymptotics for $q(s)$ and $F_2(s)$. 
From \eqref{qq} one easily obtains (correcting a misprint in the last term in Ref. \cite{BaikAsymptTW})
\be \label{qs} 
q(s) = \sqrt{\frac{-s}{2}} (1 + \frac{1}{8 s^3} - \frac{73}{128 s^6} + \frac{10657}{1024 s^9} + O(|s|^{-12})) 
\quad , \quad \psi_0(s) = \frac{s}{2} + \frac{1}{8 s^2} - \frac{9}{16 s^5} + O(\frac{1}{|s|^8})  \dots 
\ee 
and, integrating twice,
\be \label{TWas} 
\log F_2(s) = - \frac{1}{12} |s|^3 - \frac{1}{8} \log(|s|) + C + \frac{3}{2^6 |s|^3} + O(\frac{1}{|s|^3})
\ee 
where obtaining the constant $C= \frac{1}{24} \log 2 + \zeta'(-1)$ requires more sophisticated 
methods
\cite{BaikAsymptTW}.\\

Let us note that the right tail approximation of $F_2(s)$, i.e. the first order in
the expansion of the FD in powers of the Airy kernel at large positive $s$, reads
\be \label{approx} 
F_2(s) = {\rm Det}[I - P_s K_{\rm Ai} P_s] \simeq 1 - {\rm Tr} P_s K_{\rm Ai} + O(e^{- \frac{8}{3} s^{3/2} })
= 1 - \int_s^{+\infty} du \int_0^{+\infty} dv {\rm Ai}(u+v)^2 + O(e^{- \frac{8}{3} s^{3/2} })
\ee
There is a corresponding approximation $\psi_0(s)= \psi_{00}(s) + \dots$ where
one neglects the non-linear term in the KP equation, i.e. also in \eqref{psi0}, leading to
\be \label{psi00} 
-4 s \psi _{00}' -2 \psi_{00} +\psi''' _{00} = 0 
\ee
which is solved as $\psi_0(s)=-q_0(s)^2$ with $q_0''=s q_0$, solved
as $q_0(s)=- {\rm Ai}(s)$, leading to \eqref{approx}. The approximation 
\eqref{approx} corresponds to the large time limit of the contribution 
of the single string states in the Bethe ansatz. Hence the latter 
obeys the linear part of the KP equation, as discussed below (see also Appendix \ref{app:bethe}).\\

We write now the equation satisfied by the subleading term in \eqref{typtyp}.
It is more
convenient to give the result for the function $\hat \psi_1$ defined in \eqref{hatpsi1}
(with $\psi_1(s)=\hat \psi'_1(s)$). One
finds that it must satisfy the following linear equation
\be \label{subdo} 
2(1- 6 a) \hat \psi_1(s) - 4 (s+3 q(s)^2) \hat \psi'_1(s) + \hat \psi_1'''(s) = 0
\ee
%\be
%-2 \psi _1(s) \left(6 a-6 \psi _0'(s)+1\right)-4
%   \left(s-3 \psi _0(s)\right) \psi _1'(s)+\psi'''_1(s) = 0
%\ee 
%which can be written also as ** recheck ** 
%\be
%\psi_1''' + 2 (1- 6 a) \psi_1 - 4 ((s+ 3 q^2) \psi_1)' = 0
%\ee 
At large $s$ it implies that $\hat \psi_1(s) \sim (-s)^{6 a-1}$. 
It is not so easy to solve this equation. However, the analysis of the Fredholm determinant
was carried in Ref. \cite{DeanFermions} and it was found that (see Appendix B there) 
\be
\hat G(t,r=s t^{1/3}) \simeq F_2(s) + t^{-2/3} \frac{\pi^2}{6} F''_2(s) + O(t^{-4/3}) 
\ee 
This predicts that the exponent $a=2/3$ and that a solution of \eqref{subdo} should
be 
\be
\hat \psi_1(s) = \frac{\pi^2}{6} \partial_s \frac{F_2''(s)}{F_2(s)} 
= \frac{\pi^2}{6} \partial_s (\psi_0(s) + \hat \psi_0(s)^2) = \frac{\pi^2}{6} ( \frac{|s|^3}{4}-\frac{7}{16}-\frac{27}{64
   |s|^3}-\frac{855}{128
   |s|^6} + O(|s|^{-9}) )
\ee
Although we have not been able to show it directly, we carried a series expansion
at large negative $s$, using \eqref{qs} to a much higher order, 
which indeed indicates that this is the case.\\

{\bf Remark 6.} We know from Refs. \cite{we,dotsenko,spohnKPZEdge,corwinDP} that the
finite time analog to \eqref{approx} is (keeping only the first order in
the expansion of the FD in traces of the finite time kernel)
\bea
&& \log \hat G(t,r) = \log {\rm Det}[ I - \sigma_{t,r} K_{\rm Ai}] = - {\rm Tr}[\sigma_{t,r} K_{\rm Ai}] + \dots =
 - \int_{r/t^{1/3}}^{+\infty} dv \int_{-\infty}^{+\infty} du 
\frac{{\rm Ai}(u+v)^2}{1 + e^{- t^{1/3} u} } + \dots \\
&& \sigma_{t,r}(u,u') = 
\frac{1}{1+ e^{r - t^{1/3} u}} \delta(u-u') 
%
%&& = \log( 1 - \int_{0}^{+\infty} dv \int_{-\infty}^{+\infty} du 
%\frac{{\rm Ai}(u+v)^2}{1 + e^{r - t^{1/3} u} } + \dots )
%= - \int_{0}^{+\infty} dv \int_{-\infty}^{+\infty} du 
%\frac{{\rm Ai}(u+v)^2}{1 + e^{r - t^{1/3} u} } \\
%&& 
%= - \int_{r/t^{1/3}}^{+\infty} dv \int_{-\infty}^{+\infty} du 
%\frac{{\rm Ai}(u+v)^2}{1 + e^{- t^{1/3} u} } + \dots
\eea 
It is easy to check that the (single trace) leading term
\bea \label{1t}
\hat \psi_1(t,r) = \partial_r \log( \hat G(t,r))|_{\rm 1 trace} = 
t^{-1/3} \int_{-\infty}^{+\infty} du 
\frac{{\rm Ai}(u+\frac{r}{t^{1/3}})^2}{1 + e^{- t^{1/3} u} } 
\eea 
is solution of the linear part of the integrated reduced KP equation \eqref{KP22}
\be \label{KP22lin} 
 \partial_t \hat \psi_1+ \frac{1}{12} \partial_r^3 \hat \psi_1  +  \frac{1}{2 t}  \hat \psi_1 =0
\ee 
Indeed one can write formally (upon expansion of the "Fermi factor" in \eqref{1t} and using the Airy propagator identity)
\be
\hat \psi_1(t,r) = \sum_{n \geq 1} a_n \frac{1}{\sqrt{t}} e^{-n r + \frac{n^3 t}{12}}   \quad 
, \quad a_n = \frac{(-1)^n}{\sqrt{4 \pi n}}
\ee
It can also be written as
\be \label{psi11} 
\hat \psi_1(t,r) = \frac{1}{\sqrt{4 \pi t}} e^{- t \partial_r^3} {\rm Li}_{1/2}(- e^{-r}) 
\ee 
It is reminiscent but {\it different} from the exact expression of the first cumulant, see formula (111) in \cite{ShortTimeSystematic} (which is the same formula with ${\rm Li}_{3/2}$ instead of
${\rm Li}_{1/2}$). The first cumulant for $\log \hat G$ has the expression ${\rm Tr} [\log(1-\sigma_{t,r}) K_{\rm Ai}]$ 
different indeed from $- {\rm Tr}[\sigma_{t,r} K_{\rm Ai}]$ above. It provides the correct small time limit $\hat \psi(r,t) \simeq p_0(r)/\sqrt{t}$ (see \eqref{serpsi} and \eqref{init}) and does also satisfy the linearised version of KP (which can be checked by direct expansion as above, or see Section \ref{subsec:hb}). Hence, although $\hat \psi_1$ satisfies linear KP, it does not have the correct initial condition.

\section{Other initial conditions: Half-Brownian and Brownian}
\label{sec:half} 

We now turn to the half-Brownian and Brownian initial conditions. We perform some checks
from the known expressions for the cumulants of $Z(x,t)$, which hint at a general mechanism 
for the KP equation to hold. It agrees with the statement of Ref. \cite{QRKP} for the half-Brownian,
and leads us to conjecture that the KP equation is also obeyed for the full Brownian. 
In the second part, we briefly study the small time large deviations for both cases,
which confirms the conjecture.\\

It was stated in Ref. \cite{QRKP} that for the half-Brownian initial condition ($B(x)$ is a one-sided Brownian
with $B(0)=0$) 
\be
Z(x,0)= e^{h(x,0)} = e^{B(x) - w x} \theta(x)
\ee 
the function $\phi(x,t,r)= \partial_r^2 \log G(x,t,r)$ associated to the same generating function 
$G$ defined in \eqref{gen0},
satisfies the KP equation, which we recall here
\be \label{KP1n} 
\partial_t \phi + \phi \, \partial_r \phi + \frac{1}{12} \partial_r^3 \phi + \partial_r^{-1} \partial_x^2 \phi
= 0
\ee
In addition we conjecture here that a similar property holds for the Brownian
initial condition
\be
Z(x,0)= e^{h(x,0)} = e^{B_R(x) - w_R x} \theta(x) + e^{B_L(x) + w_L x} \theta(-x) 
\ee 
for the modified generating function
\be \label{gen0} 
G(x,t,r) = \big\langle \exp(- e^{ h(x,t) + \chi + \frac{t}{12} - r}) \big\rangle 
= \big\langle \exp(- \tilde Z(x,t) e^{\frac{t}{12} - r}) \big\rangle
\ee 
where $\chi$ is a log-gamma random variable independent from $h(x,t)$, of PDF
$P(\chi)d\chi=e^{-2 w \chi - e^{-\chi}} d\chi/\Gamma(2w)$ and of exponential moments 
$e^{n \chi} = \frac{\Gamma(w - n)}{\Gamma(w)}$, with $2 w=w_R+w_L$. 
We claim that $G$ also satisfies the KP equation \eqref{KP1n}.\\

We can again expand $G$ in \eqref{gen0} in cumulants, now in presence of space dependence.
This leads to the series in $e^{-r}$
\be
\phi(x,t,r)= \sum_{n \geq 1} \frac{(-1)^n n^2}{n!} Z_n(x,t) e^{\frac{n t}{12} -n r}  \quad , \quad 
Z_n(x,t) = \langle Z(x,t)^n \rangle_c 
\ee
In the case of the full Brownian, $Z(x,t)$ must be replaced by $\tilde Z(x,t)= e^{h(x,t) + \chi}$
and the cumulants are averages over both $h$ and $\chi$. 
Inserting in the equation and identifying the terms $e^{- n r}$ we obtain the recursion, for $n \geq 1$
\bea \label{genN} 
- \partial_t Z_n(x,t) 
+ \frac{n^3-n}{12} Z_n(x,t) 
+ \frac{1}{n} \partial_x^2 Z_n(x,t)  = 
- \frac{1}{2} (n-1)! \sum_{n_1+n_2=n, n_1,n_2 \geq 1}
\frac{n_1^2 n_2^2}{n_1! n_2!} 
Z_{n_1}(x,t) Z_{n_2}(x,t) 
\eea 
which, as in the previous Section, 
allows to determine the moments recursively from the first one, up to
some undetermined solution of the homogeneous equation which enters at each level $n$.
The latter reads
\bea \label{homo}
z_n(x,t)= e^{ \frac{n^3-n}{12} t} \int \frac{dk}{2 \pi} \hat z_n(k) e^{- n t k^2 + i n k x}
\eea
where $\hat z_1(k)$ is the Fourier transform of $Z_1(x,0)$ (which is
usually specified by the initial data) and the 
$\hat z_n(k)$ are (up to a rescaling) the Fourier transforms of the $z_n(x,0)$ 
are a priori arbitrary (if we consider the general solution of KP). As a side remark,
see Appendix \ref{app:bethe} for details and definitions, 
Eq. \eqref{homo} corresponds to the evolution of the {\it moments} 
$\langle Z(x,t)^n \rangle_{\rm 1 string}$ retaining only the
contribution of the eigenstates of the delta bose-gas Hamiltonian 
corresponding to a single
string $n_s=1$ with arbitrary momentum (which contains the
ground state). The general solution of the KP equation is thus a functional of the set
of functions $\hat z_j(k)$
\be
Z_n(x,t) = F_n(x,t ; \{ \hat z_j(k) \}_{j=1,\dots,n})
\ee
and it is a priori far from obvious in general that this corresponds to the exact cumulants
of the KPZ/SHE equation for {\it some} initial condition. Let us now make some more detailed analysis.\\

Let us start with $n=1$, which reads
\bea \label{eqrec1} 
- \partial_t Z_1(x,t) + \partial_x^2 Z_1(x,t)  = 0
\eea 
Hence $Z_1(x,t)=z_1(x,t)$ must satisfy the heat equation, which as we discussed above is
a consequence of the Ito convention. Hence at this stage any initial condition of the SHE would work.
Indeed we see that for the half-Brownian $Z_1(x,t)=\langle Z(x,t) \rangle$ given in
\eqref{z1} does satisfy \eqref{eqrec1}. Although it does involve some additional
averaging over the Brownian IC, the linearity of \eqref{eqrec1} guarantees that it works.\\

Let us write now the equation for the second cumulant $Z_2(x,t)$
\bea \label{cumx2} 
\partial_t Z_2(x,t) 
- \frac{1}{2} Z_2(x,t) 
- \frac{1}{2} \partial_x^2 Z_2(x,t)  = 
\frac{1}{2} Z_{1}(x,t)^2
\eea 
The general solution is
\bea
Z_2(x,t)  = z_2(x,t) + 
\frac{1}{2}  \int dx' \int_0^t dt' \frac{1}{\sqrt{2 \pi(t-t')}} e^{- \frac{(x-x')^2}{2 (t-t')} }
Z_{1}(x',t')^2 e^{\frac{1}{2} (t-t')} 
\eea 
which is not illuminating. In the Appendix \ref{app:bethe} the following form, suggested by the Bethe ansatz, is studied. 
Suppose that the generic integer
moment can be written as 
\bea \label{zbethe} 
&& \langle Z(x,t)^n \rangle = n! \sum_{n_s=1}^n \frac{1}{n_s!} \sum_{(m_1,\dots,m_{n_s})_n}  \prod_{j=1}^{n_s} \int \frac{dk_j}{2 \pi m_j}  e^{i x \sum_{j=1}^{n_s} m_j k_j} e^{t \sum_{j=1}^{n_s} [ \frac{1}{12} (m_j^3-m_j) - m_j k_j^2]}
\\
&& \times \prod_{j=1}^{n_s}  S_{k_j,m_j} 
\prod_{1 \leq i < j \leq n_s} \frac{4(k_i-k_j)^2 + (m_i-m_j)^2}{4(k_i-k_j)^2 + (m_i+m_j)^2} \nn
\eea 
which is the case for the half-Brownian with $S_{k,m}= \frac{\Gamma \left(w+i k-\frac{m}{2}\right)}{\Gamma
   \left(w+i k+\frac{m}{2}\right)}$, and for the
   Brownian (for the modified partition sum $\tilde Z(x,t)$) with 
   $S_{k,m}= \frac{\Gamma \left(w_R+i k-\frac{m}{2}\right)}{\Gamma
   \left(w_R+i k+\frac{m}{2}\right)}  \frac{\Gamma \left(w_L+i k-\frac{m}{2}\right)}{\Gamma
   \left(w_L+i k+\frac{m}{2}\right)}$, an information obtained from
   the replica Bethe ansatz solutions 
   of Ref. \cite{SasamotoHalfBrownReplica,SasamotoStationary,SasamotoStationary2}.
   Then, irrespective of the precise form of $S_{k,m}$ the
  equations \eqref{genN} will be obeyed. This conjecture is verified in the Appendix \ref{app:bethe}
  for $n=2,3$. A similar mechanism holds for the nested contour integral form
  (see Appendix).\\
  
Hence we expect that the KP equation will be obeyed for any IC such that the overlap factorizes. 
Let us now examine the question of the initial condition, and the
short time large deviations. 

\subsection{Half-Brownian} 
\label{subsec:hb} 

For the half-Brownian IC, the initial data for the moments read
\be
\langle Z(x,0)^n \rangle = \langle e^{n B(x)}  \rangle_B e^{-n w x} \theta(x) 
= e^{n (\frac{n}{2} - w) x} \theta(x)
\ee 
and one can also write explicitly the initial condition for $G$ (which is discontinuous at $x=0$)
\be
G(x,0,r) = \langle e^{- \theta(x) e^{B(x)} e^{-w x -r}} \rangle
= \int_{-\infty}^{+\infty} \frac{db}{\sqrt{2 \pi x}} e^{- \frac{b^2}{2 x} - e^{b - w x - r}} \theta(x) + \theta(-x)
\ee 

We now study the time evolution at short time $t \ll 1$. The regime of interest
will correspond to small $x$, large $w$ and $x w$ fixed, so we will need
the initial condition in that region. Let us write explicitly the initial condition for the first three cumulants and their expansion at small $x$, large $w$ and $x w$ fixed
\bea
&& Z_1(x,t=0)= e^{(\frac{1}{2} - w) x} \theta(x) \simeq e^{- w x} \theta(x) \\
&& Z_2(x,t=0)= (e^{2 (1- w) x} - e^{2 (\frac{1}{2} - w) x}) \theta(x) \simeq x e^{- 2 w x}  \theta(x)  \\
&& Z_3(x,t=0)=\left(e^x-1\right)^2 \left(e^x+2\right)
   e^{\left(\frac{3}{2}-3 w\right) x} \theta(x) \simeq 3 x^2 e^{- 3 w x}  \theta(x)
\eea
It is easy to guess that the general formula is 
\be \label{incond} 
Z_n(x,t=0) \simeq n^{n-2} x^{n-1} e^{- n w x}  \theta(x) 
\ee 
We will need the definition
and series expansion of the standard branch $W_0(z)$ of the Lambert function \cite{KnuthLambert},
$W(z)$ 
\be \label{Lambert} 
W(z) e^{W(z)} = z \quad , \quad 
W(z)=W_0(z)=- \sum_{n\geq 1} \frac{(-1)^n}{n!} n^{n-1} z^n
\ee
From the moments we can thus write the initial condition for $G$, in that regime
(for small $x$, large $w$, with $w x$ fixed) as 
\be
\log G(x,t=0,r) = \sum_{n \geq 1} \frac{(-1)^n}{n!} Z_n(x,t=0) e^{-n r} 
\simeq
\frac{1}{x} \sum_{n \geq 1} \frac{(-1)^n}{n!} n^{n-2} e^{- n (r - \log(x) + w x)} \theta(x) 
\ee 
and its derivative, 
\be \label{Gt0} 
\hat \phi(x,t=0,r) = \partial_r \log G(x,t=0,r) =\frac{1}{x} W_0(e^{-(r-\log x + w x)} )  \theta(x) 
\ee
For fixed $r$ and small $x$ with $w x$ fixed one has $\partial_r \log G(x,t=0,r) \simeq e^{-r - w x} \theta(x)$ 
as for a half-wedge, however there is a fluctuation region, with fixed $r-\log x + w x$ where 
$\partial_r  \log G(x,t=0,r) \sim 1/x $ is large.\\

To study the small time large deviations of small $t$, small $x$ and large $w$ 
with $w x$ fixed we define 
\be
\tilde x = \frac{x}{\sqrt{t}}  \quad , \quad \tilde w = w \sqrt{t} 
\ee
By analogy with the study of the Brownian IC in Ref. \cite{krajenbrink2017exact}, we expect in the
case of the half-Brownian, the large deviation form at $\tilde x=0$
\be
\log G(0,t,r) \simeq - \frac{1}{\sqrt{t}} \Psi( t^{1/2} e^{-r}) 
\ee 
At finite $\tilde x$ we thus expect the following form (which we obtain explicitly below)
\be
 \log G(x,t,r) \simeq - \frac{1}{\sqrt{t}} \hat \Psi(\tilde x, r - \frac{1}{2} \log t) \quad , \quad  \phi(x,t,r) 
 = \partial_r^2 \log G(x,t,r) \simeq - \frac{1}{\sqrt{t}} \psi(\tilde x, r - \frac{1}{2} \log t) 
\ee
with $\psi=\partial_r^2 \hat \Psi$. 
Inserting into the KP equation \eqref{KP1} gives to leading order $O(t^{-3/2})$
\be \label{lin2} 
\partial_r \psi + \partial_r^2 \psi + x \partial_r \partial_x \psi - 2 \partial_x^2 \psi = 0
\ee 
Note that for the droplet IC the same equation degenerates and does not allow to 
determine the function. Since \eqref{lin2} is a linear equation, an alternate method, which we now use,
is to study the linear part of the recursion \eqref{genN} for the cumulants
\bea \label{genN2} 
 [ \partial_t - \frac{n^3-n}{12} -  \frac{1}{n} \partial_x^2 ] Z_n(x,t)  = 0
\eea 
At short time one can neglect the term $(n^3-n)/12$, and the solution can be written
explicitly from the knowledge of the initial condition \eqref{incond}, as
\be \label{solut} 
Z_n(x,t) \simeq n^{n-2} \int_0^{+\infty} \frac{dy}{\sqrt{4 \pi t/n}} e^{- \frac{n (x-y)^2}{4  t} } y^{n-1} e^{-n w y}
\ee 
Other, more explicit, expressions for these cumulants are given in Appendix \ref{app:cum}.
\\

Before evaluating this expression, let us first show that at large $w$ one recovers exactly the droplet result.
The large $w$ limit is obtained setting $y \to y/w$ which leads to
\be
Z_n(x,t) \simeq \frac{1}{\sqrt{4 \pi t}} n^{n-\frac{3}{2}} w^{-n} e^{- \frac{n x^2}{4  t} }
\int_0^{+\infty} dy  y^{n-1} e^{-n  y} = 
\frac{1}{\sqrt{4 \pi t}} n^{-\frac{3}{2}} \Gamma(n) w^{-n} e^{- \frac{n x^2}{4  t} }
\ee 
and to 
\be
\log G(x,t,r) \simeq \frac{1}{\sqrt{4 \pi t}} \sum_{n \geq 1} (-1)^n  n^{-5/2} (\frac{1}{w} e^{-r - \frac{x^2}{4 t}} )^n
= \frac{1}{\sqrt{4 \pi t}} {\rm Li}_{\frac{5}{2}}( - \frac{1}{w} e^{-r - \log w - \frac{x^2}{4 t}}  )
%= \frac{1}{\sqrt{t}} \Psi(- e^{-r - \log w - \frac{x^2}{4 t}} )
\ee 
Apart from a shift, this is exactly the result obtained in Ref. \cite{le2016exact}. The shift is
easy to understand. Indeed, since at large $w$, $e^{-w x+B(x)} \theta(x) \to \frac{1}{w} \delta(x)$,
we expect $Z_{hb}(x,t) \to \frac{1}{w} Z_d(x,t)$ (the index hb refers to half-Brownian and $d$ to
droplet), that is $\langle \exp(-e^{h_{hb}(x,t)-r}) \rangle \to 
\langle \exp(-e^{h_{d}(0,t)-r-\log w -\frac{x^2}{4t}}) \rangle = 
\langle \exp(-e^{h_{d}(x,t)-r-\log w}) \rangle$. Hence using the half-Brownian IC as
a regularisation to run the KP equation, one can indeed calculate the leading term
$p'_0(r)={\cal L}_1$ in the short time expansion for the droplet IC (which was
missing from a direct approach). \\

Let us return to the solution \eqref{solut} for the cumulants and perform the summation
(one can neglect the term $t/12$ in the exponential at small time)
\bea 
&& \log G(x,t,r) \simeq \sum_{n \geq 1} \frac{(-1)^n}{n!} Z_n(x,t) e^{- n r} 
 \simeq \frac{1}{\sqrt{4 \pi t}} \int_0^{+\infty} \frac{dy}{y} \sum_{n \geq 1} \frac{(-1)^n}{n!} n^{n- \frac{3}{2}}
[y e^{- r - \frac{(x-y)^2}{4 t} - w y} ]^n \\
&& = \frac{1}{\sqrt{4 \pi t}} \int_0^{+\infty} \frac{dy}{y} \sum_{n \geq 1} \frac{(-1)^n}{n!} n^{n- \frac{3}{2}}
[y e^{- r + \frac{1}{2} \log t - \frac{(\tilde x-y)^2}{4} - \tilde w y } ]^n \label{ser3} 
\eea 
where in the second expression we have rescaled $y \to \sqrt{t} y$. We use now the
following integral representation for the series
\bea
&& \sum_{n \geq 1} \frac{(-1)^n}{n!} n^{n- \frac{3}{2}} z^n 
%= 
%- \sum_{n \geq 1} \frac{(-1)^{n-1}}{n!} n^{n-1} \frac{1}{\sqrt{n}} z^n
= - \int_0^{+\infty} \frac{du}{\sqrt{\pi u}} \sum_{n \geq 1} \frac{(-1)^{n-1}}{n!} n^{n-1} (z e^{-u}) ^n 
 = - \int_0^{+\infty} \frac{du}{\sqrt{\pi u}}  W_0(z e^{-u}) 
\eea
using the formula \eqref{Lambert} for the Lambert function $W_0$.
Hence we find our final result for the small time large deviation function 
from the Brownian IC 
\bea \label{explicit2} 
&& \log G(x,t,r) \simeq - \frac{1}{\sqrt{t}} \hat \Psi_{\tilde w}(\tilde x, r - \frac{1}{2} \log t) \quad , \quad \tilde x=x/\sqrt{t} 
\quad , \quad \tilde w = w \sqrt{t} \\
&& \hat \Psi_{\tilde w}(\tilde x, r) = \frac{1}{\sqrt{4 \pi}} \int_0^{+\infty} \frac{dy}{y} 
\int_0^{+\infty} \frac{du}{\sqrt{\pi u}}  W_0(y e^{- r - u - \frac{(\tilde x-y)^2}{4} - \tilde w y}) \nn
\eea which gives Eq. \eqref{Ppsi2} in the introduction, with $\Psi(\tilde x, z=e^{-r}) = \hat \Psi(\tilde x,r)$.\\

This result can be put in an equivalent form \cite{footnoteK}. Going back to the series \eqref{ser3} we can rescale $y \to y/n$, expand the square in the exponential,  
and use the representation $e^{- \frac{y^2}{4 n}} = \sqrt{\frac{n}{\pi}}  \int_{-\infty}^{+\infty} dk 
e^{i k y - n k^2}$ to put it in the form
\be
\log G(x,t,r) \simeq \frac{1}{\sqrt{t}} \int_{-\infty}^{+\infty} \frac{dk}{2 \pi} 
\sum_{n \geq 1} \frac{(-1)^n}{n n!} (e^{- r - \frac{\tilde x^2}{4} + \frac{1}{2} \log t - k^2})^n 
\int_0^{+\infty} \frac{dy}{y} y^n e^{- y ( \tilde w - \frac{\tilde x}{2} - i k)} 
\ee 
When $\tilde w - \frac{\tilde x}{2} >0$ one can perform the integral over $y$, and one recognizes the series
expansion of the function ${\rm Li}_2(s)=\sum_{n \geq 1} \frac{s^n}{n^2}$, leading to 
\be \label{alternate} 
\log G(x,t,r) \simeq \frac{1}{\sqrt{t}} \int_{-\infty}^{+\infty} \frac{dk}{2 \pi} 
{\rm Li}_2( - \frac{e^{- r - \frac{\tilde x^2}{4} + \frac{1}{2} \log t } e^{-k^2} }{ \tilde w - \frac{\tilde x}{2} - i k}) 
\ee 
This form of the large deviation function is similar to the generic form which is obtained for other
solvable cases \cite{K2}.

\subsection{Brownian}

We now consider briefly the full Brownian IC. The initial condition is
\be
\tilde Z(x,0) = e^{\chi} ( e^{B_R(x)} e^{-w_R x} \theta(x) 
+  e^{B_L(x)} e^{w_L x} \theta(-x) )
\ee 
and its moments are given by (we denote $2 w=w_L+w_R$)
\be
 \langle \tilde Z(x,0)^n \rangle 
%= <e^{n \chi}> \bigg(  \langle e^{n B_R(x)}  \rangle_{B_R} e^{-n w_R x} \theta(x) 
%+  \langle e^{n B_L(x)}  \rangle_{B_L} e^{n w_L x} \theta(-x) \bigg)
%\\
 =  \frac{\Gamma(2 w - n)}{\Gamma(2 w)} \bigg( e^{n (\frac{n}{2} - w_R) x} \theta(x) + e^{n (\frac{n}{2} + w_L) x} \theta(-x) \bigg)
\ee 
The corresponding initial condition for $G$ can be written as
\be
G(x,0,r) = \langle e^{- \tilde Z(x,0) e^{ -r}} \rangle
= \int_{-\infty}^{+\infty} d\chi P(\chi) 
\int_{-\infty}^{+\infty} \frac{db}{\sqrt{2 \pi |x|}} e^{- \frac{b^2}{2 |x|} }
(e^{- e^{b - w_R x + \chi - r}} \theta(x) + e^{- e^{b + w_L x + \chi - r}} 
\theta(-x) )
\ee 

Let us specify to the case $w_L=w_R=w$. As for the half-Brownian, we want to study the limit 
of small $x$, $w$ large with $w x$ fixed. We would like to apply the same method as for
the half-Brownian, i.e. evolve the cumulants with the linear part of the KP equation as
\be \label{evo} 
\tilde Z_n(x,t) \simeq  \int_{-\infty}^{+\infty} \frac{dy}{\sqrt{4 \pi t/n}} e^{- \frac{n (x-y)^2}{4  t} } \tilde Z_n(y,t=0)
\ee 
However the
initial condition for
$\tilde Z_n(x,t=0)$ in the regime of interest is now more complicated, e.g.
\bea \label{ico}
&& \tilde Z_1(x,t=0) \simeq \frac{1}{2 w} e^{- w |x|} \quad , \quad  \tilde Z_2(x,t=0) \simeq \frac{1}{8 w^3} (1 + 2 w |x|) e^{- 2 w |x|} \\
&& \tilde Z_3(x,t=0) \simeq \frac{1}{8 w^5} (1 + 3 w |x|(1+ w|x|)) e^{- 3 w |x|} 
\quad , \quad \tilde Z_4(0,t=0) \simeq \frac{15}{64 w^5}
\eea 
and so on, with more and more complicated polynomials. Hence we have performed only a few checks on the following
conjectured formula
\bea
&& \log G(x,t) = \sum_{n \geq 1} \frac{(-1)^n}{n!} \tilde Z_n(x,t) e^{-n r}|_{x=\sqrt{t} \tilde x, 
w=\tilde w/\sqrt{t}} |_{\tilde x=0} \simeq - \frac{1}{\sqrt{t}} \Psi(t e^{-r}) \\
&& \Psi(z) = \frac{1}{\pi} \int_0^{+\infty} dy (1+ \frac{1}{y+\tilde w^2}) \sqrt{y} \log(1 + \frac{z e^{-y}}{y+\tilde w^2}) = - \int_{-\infty}^{+\infty} \frac{dk}{2 \pi} {\rm Li}_2(- z \frac{e^{-k^2}}{k^2 + \tilde w^2}) 
\eea 
where the second line is the exact result for the Brownian IC obtained from the FD in Ref. \cite{krajenbrink2017exact} (the equivalent last expression was obtained 
in \cite{ShortTimeSystematic} and note that all known
solvable IC for KPZ in full space and half-space can be put in similar forms \cite{AlexKrajPHD}).
Inserting the $\tilde Z_n(x,t)$ obtained from \eqref{evo} and the initial condition \eqref{ico} we have verified by
series expansion in $e^{-r}$ that it holds for $n=1,2,3$. Although much remains to be done,
this provides a nice check that the full Brownian IC indeed satisfies the KP equation, as claimed here.

\section{Conclusion} 

In conclusion we have studied some of the consequences of the property recently discovered 
in Ref. \cite{QRKP},
that the generating function of the droplet and half-Brownian IC solutions of the KPZ equation 
satisfy the KP equation. We have also studied the mechanism for this property to hold
on the cumulants $Z_n(x,t)$, which led to the conclusion that the modified generating function for the full Brownian IC (or any IC with a "decoupled" overlap) should also
satisfy this property.\\

The main consequences of the KP property studied here concerns the large deviations, both at short time and at large time. For the droplet IC we have found that the question of which initial condition should be used for the KP equation is intimately related to the small time large deviations. In the case of the droplet IC, the KP equation simplifies into a reduced KP equation. We have shown how to recover, from this reduced KP equation, and in a rather effortless way, the full systematic short time expansion obtained previously in Ref. \cite{ShortTimeSystematic}. On the other hand substituting the large time large deviation form in the reduced KP equation provides a (rather simple)
fifth method to obtain the rate function $\Phi_-(z)$, up to a single undetermined parameter. We showed how
this parameter can be determined using the so-called cumulant summation of the short time expansion.
This method, which allows to relate the short time and the large time large deviation regimes, was studied within the KP setting. It takes the form of a semi-classical expansion where one treats the 
third derivative term in the KP equation as a perturbation. It can nicely be solved in terms of Burgers equation. This allowed us also to obtain in addition the first subdominant corrections, not obtained in Ref. \cite{ShortTimeSystematic}. We have shown that not only the KPZ problem, but a variant of a more general
problem of linear statistics of the Airy point process, defined and studied in Ref. \cite{ShortTimeSystematic} and \cite{largedev}, do obey the KP property. We have re-obtained some results for these linear statistics, by a completely different method. The ensuing connections between the KP equation, the Coulomb gas, the non-local Painleve equation, and the stochastic Airy operator (the connections between the latter three were unveiled in Ref. \cite{largedev})
remain to be investigated deeper. \\

Note two interesting consequences. The first arises from the connection 
\cite{FermionsT,DeanFermions}
of the droplet solution of the KPZ equation to $N$ non interacting fermions in a 1D harmonic trap at finite temperature $T$, with Hamiltonian $H= \frac{p^2}{2} + \frac{x^2}{2}$, in the limit $N,T \to +\infty$ at
fixed $b=N^{1/3}/T$. Denoting $x_{\rm max}(T)$ the position of the rightmost fermion, and $\xi=\sqrt{2} N^{1/6} (x_{\rm max}(T)-\sqrt{2 N})$ the centered
and scaled position, then $\partial_r^2 \log {\rm Prob}(b \xi < r)$ satisfies the reduced KP 
equation \eqref{KP2} with $t=b^3$. A second consequence arises from the connection between the KPZ equation and the non-relativistic limit of the D=1+1 sine Gordon field theory \cite{SineGordon}.
It implies that, in that limit, the two time correlation function of the field $e^{a \varphi(0,t)}$ 
(at time zero and $t$, $\varphi$ being the sine Gordon field) identifies with the generating function $\hat G(t,r)$ studied in the present work, with the correspondence $e^{\frac{t}{12} - r} =(2 \sinh \frac{a}{2})^2 e^{-M c_\ell^2 t}$.
Hence, in that limit the two time correlation function obeys a differential equation
related to KP via this change of variable.\\

For the half-Brownian and Brownian IC the study is technically more involved since one should handle 
space and time, and many questions remain. However, we have obtained the small time large deviation space-time rate function for the half-Brownian IC. We have also checked that the formal solution at short time in the case of the Brownian IC does agree with known results for the large deviation rate function at the origin, from Ref. \cite{krajenbrink2017exact}.\\

Although the present study does not explain the deep reason of why the KP equation appears
in some finite time solutions to the KPZ equation (which is related to why they can be expressed as a specific form of Fredholm determinants) it does show that this property provides a new interesting angle
to study properties of these solutions. 

\acknowledgments
I thank J. Quastel and D. Remenik for sharing some details about their work. 
I thank G. Barraquand, M. Cafasso, S. N. Majumdar, G. Mussardo, S. Prolhac and G. Schehr for discussions. 
I am particularly grateful to A. Krajenbrink for an ongoing collaboration, and discussions on related topics. 
I acknowledge support from ANR grant ANR-17-CE30-0027- 01 RaMaTraF.

\appendix

\section{Short time expansion} 
\label{app:shorttime}

We give here more details on the short time expansion for droplet IC. It is also valid for
more general linear statistics problems see Remark 2. \\

Inserting the series \eqref{serpsi} into \eqref{KP2} and integrating once over $r$ (as in \eqref{KP22})
we obtain the recursion, for $m \geq 0$
\bea
\frac{m+1}{2} p_{m+1} + \frac{1}{12} \theta_{m \geq 1} p'''_{m-1} + \frac{1}{2} \sum_{m_1+m_2=m, m_1 \geq 0, m_2 \geq 0} p'_{m_1} p'_{m_2} = 0
\eea 
%Hence the recursion
%\bea
%&& p_1 = - (p'_0)^2 \\
%&& p_{m+1} = - \frac{2}{m+1} \left( \frac{1}{12} p'''_{m-1} 
%+  \sum_{j=0}^{\frac{m-1}{2}} p'_{j} p'_{m-j}  \right) \quad , \quad \text{m odd} \ge 1 \\
%&& p_{m+1} = - \frac{2}{m+1} \left( \frac{1}{12} p'''_{m-1} 
%+  \sum_{j=0}^{\frac{m}{2}-1} p'_{j} p'_{m-j} +
%\frac{1}{2} (p'_{m/2})^2  \right) \quad , \quad \text{m even} \ge 2
%\eea 
Performing the recursion with Mathematica we find that all $p_m$ are total 
derivatives of the form $p_m= \partial_r P_m(p_0',p_0'',..)$ where $P_m$ are
polynomials. Using $- \partial_r {\cal L}_i = {\cal L}_{i+1}$ and 
$p'_0(r) = {\cal L}_1$, we find exactly all seven terms (up to $O(t^3)$) given in the lengthy equation (61)
in Ref. \cite{ShortTimeSystematic}. It is easy to obtain quickly the next terms, and we will show
here only the next one $O(t^{7/2})$
\bea
&& p_8(r) = \partial_r \bigg[ \frac{2}{315} {\cal L}_7 {\cal L}_1^8+\frac{32}{63} {\cal L}_4^2
   {\cal L}_1^7+\frac{16}{21} {\cal L}_3 {\cal L}_5 {\cal L}_1^7+\frac{32}{105}
   {\cal L}_2 {\cal L}_6 {\cal L}_1^7+\frac{16}{3} {\cal L}_3^3 {\cal L}_1^6+\frac{64}{3}
   {\cal L}_2 {\cal L}_3 {\cal L}_4 {\cal L}_1^6+\frac{16}{3} {\cal L}_2^2 {\cal L}_5
   {\cal L}_1^6+\frac{1}{135} {\cal L}_8 {\cal L}_1^6  \nn \\
   && +96 {\cal L}_2^2 {\cal L}_3^2
   {\cal L}_1^5 
    +\frac{128}{3} {\cal L}_2^3 {\cal L}_4 {\cal L}_1^5+\frac{10}{9}
   {\cal L}_4 {\cal L}_5 {\cal L}_1^5+\frac{32}{45} {\cal L}_3 {\cal L}_6
   {\cal L}_1^5+\frac{4}{15} {\cal L}_2 {\cal L}_7 {\cal L}_1^5+\frac{28}{3} {\cal L}_2
   {\cal L}_4^2 {\cal L}_1^4+160 {\cal L}_2^4 {\cal L}_3 {\cal L}_1^4+\frac{118}{9} {\cal L}_3^2
   {\cal L}_4 {\cal L}_1^4 \nn \\ 
   &&   + \frac{44}{3} {\cal L}_2 {\cal L}_3 {\cal L}_5   {\cal L}_1^4 
   +\frac{10}{3} {\cal L}_2^2 {\cal L}_6 {\cal L}_1^4+\frac{1}{432}
   {\cal L}_9 {\cal L}_1^4+\frac{128}{3} {\cal L}_2^6 {\cal L}_1^3+\frac{128}{3}
   {\cal L}_2 {\cal L}_3^3 {\cal L}_1^3+\frac{17}{90} {\cal L}_5^2
   {\cal L}_1^3+\frac{280}{3} {\cal L}_2^2 {\cal L}_3 {\cal L}_4
   {\cal L}_1^3+\frac{160}{9} {\cal L}_2^3 {\cal L}_5 {\cal L}_1^3 \nn \\
   && +\frac{83}{270}
   {\cal L}_4 {\cal L}_6 {\cal L}_1^3  +\frac{89}{540} {\cal L}_3 {\cal L}_7
   {\cal L}_1^3+\frac{1}{18} {\cal L}_2 {\cal L}_8 {\cal L}_1^3+\frac{320}{3}
   {\cal L}_2^3 {\cal L}_3^2 {\cal L}_1^2+\frac{109}{36} {\cal L}_3 {\cal L}_4^2 {\cal L}_1^2+40
   {\cal L}_2^4 {\cal L}_4 {\cal L}_1^2+\frac{7}{3} {\cal L}_3^2 {\cal L}_5
   {\cal L}_1^2+\frac{17}{5} {\cal L}_2 {\cal L}_4 {\cal L}_5 {\cal L}_1^2 \nn \\
   && +\frac{32}{15}
   {\cal L}_2 {\cal L}_3 {\cal L}_6 {\cal L}_1^2  +\frac{5}{12} {\cal L}_2^2 {\cal L}_7
   {\cal L}_1^2+\frac{{\cal L}_{10} {\cal L}_1^2}{5184}+\frac{14}{9} {\cal L}_3^4
   {\cal L}_1+\frac{29}{6} {\cal L}_2^2 {\cal L}_4^2 {\cal L}_1+32 {\cal L}_2^5 {\cal L}_3
   {\cal L}_1+\frac{40}{3} {\cal L}_2 {\cal L}_3^2 {\cal L}_4 {\cal L}_1+\frac{22}{3}
   {\cal L}_2^2 {\cal L}_3 {\cal L}_5 {\cal L}_1 \nn \\
&&   +\frac{10}{9} {\cal L}_2^3 {\cal L}_6
   {\cal L}_1  +\frac{37 {\cal L}_5 {\cal L}_6 {\cal L}_1}{1512}+\frac{103 {\cal L}_4 {\cal L}_7
   {\cal L}_1}{6048}+\frac{17 {\cal L}_3 {\cal L}_8
   {\cal L}_1}{2160}+\frac{1}{432} {\cal L}_2 {\cal L}_9 {\cal L}_1+\frac{16
   {\cal L}_2^7}{21}+\frac{14}{3} {\cal L}_2^2 {\cal L}_3^3+\frac{583
   {\cal L}_4^3}{18144}+\frac{607 {\cal L}_2
   {\cal L}_5^2}{15120} \nn \\
   && +\frac{58}{9} {\cal L}_2^3 {\cal L}_3
   {\cal L}_4  +\frac{5}{6} {\cal L}_2^4 {\cal L}_5+\frac{1121 {\cal L}_3 {\cal L}_4
   {\cal L}_5}{7560}+\frac{17}{360} {\cal L}_3^2 {\cal L}_6+\frac{503 {\cal L}_2
   {\cal L}_4 {\cal L}_6}{7560}+\frac{77 {\cal L}_2 {\cal L}_3
   {\cal L}_7}{2160}+\frac{5}{864} {\cal L}_2^2
   {\cal L}_8+\frac{{\cal L}_{11}}{497664} \bigg] \nn
\eea 
where we recall that $p'_m(r)$ is also the term of order $t^{\frac{m-1}{2}}$ in
$q_{t,\beta=1}(\sigma)$ as defined in equation (61)
in Ref. \cite{ShortTimeSystematic}.

\section{Checks on cumulants from the Bethe ansatz for the droplet, half-Brownian and Brownian IC}
\label{app:bethe} 

Let us recall briefly that the moments of the solution of
the SHE can be obtained as a sum over the 
eigenstates of the delta Bose gas (Lieb Liniger) Hamiltonian \cite{ll} 
$H_n =  - \sum_{\alpha=1}^n \partial^2_{x_\alpha} - 2 \bar c \sum_{1 \leq \alpha < \beta \leq n} \delta(x_\alpha-x_\beta)$, as follows
\bea \label{sumf}
\langle Z(x,t)^n \rangle = \sum_\mu \Psi_\mu(x,\dots,x) \frac{e^{-t E_\mu}}{||\mu||^2}  \langle \Psi_\mu| \Phi_0 \rangle 
= \sum_\mu \Psi^*_\mu(x,\dots,x) \frac{e^{-t E_\mu}}{||\mu||^2}  \langle \Phi_0 | \Psi_\mu \rangle
\eea 
in quantum mechanical notations, where $\Psi_{\mu}$ denote the eigenfunctions (and $||\mu||$ their norm) and $E_\mu$ the eigenvalues of $H_n$. Note that $\bar c=1$ in our units for the study of the SHE (corresponding to attractive interactions), but the
case $\bar c=-c<0$ is also of interest in the context of repulsive bosons
\cite{Coincidences}. The eigenfunctions \cite{ll} are parameterized by a set of rapidities
$\mu \equiv \{ \lambda_1,..\lambda_n\}$, are totally symmetric in the $x_\alpha$, and in the sector $x_1 \leq x_2 \leq \dots \leq x_n$, take the (un-normalized) form of a sum over permutations $P$ 
\be \label{def1}
\Psi_\mu(x_1,..x_n) =  \sum_{P \in S_n} A_P \prod_{j=1}^n e^{i \sum_{\alpha=1}^n \lambda_{P_\alpha} x_\alpha} \, , \quad 
A_P=\prod_{1 \leq \alpha < \beta \leq n} \Big(1 + \frac{i  
%~{\rm sgn}(x_\beta - x_\alpha)
}{\lambda_{P_\beta} - \lambda_{P_\alpha}}\Big)\,.
\ee
with $E_\mu=\sum_{\alpha=1}^n \lambda_\alpha^2$. To evaluate \eqref{sumf} one needs 
$\Psi^*_\mu(x,..x) = n! e^{-  i x \sum_\alpha \lambda_\alpha}$. For the Brownian IC 
the wavefunction of the initial replica state is:
\bea
&& \Phi_0(Y) = \langle y_1,..y_n | \Phi_0 \rangle = 
\langle \prod_{\alpha=1}^n ( e^{w_L y_\alpha} e^{ B_L(- y_\alpha) }  \theta(-y_\alpha) + e^{-w_R y_\alpha}  e^{ B_R(y_\alpha) } \theta(y_\alpha) ) \rangle_{B_L,B_R}  
\eea 
where $Y \equiv y_1,..y_n$. The half-Brownian is obtained setting $w_L \to +\infty$, and for the droplet IC, 
$\Phi_0(Y)= \prod_{\alpha=1}^n \delta(y_\alpha)$, obtained e.g. by further multiplying by $w_R^n$ and
sending $w_R \to +\infty$. One needs the overlap
\bea
&& \langle \Phi_0 | \Psi_\mu \rangle = n! \int_{ y_1<y_2< ..<y_n } \Psi_\mu(Y) \Phi_0(Y) 
 = n! \sum_{P \in S_n} A_P \int_{y_1<y_2< ..<y_n } e^{i \sum_{\alpha=1}^n \lambda_{P_\alpha} y_\alpha} \Phi_0(Y)
\eea 
A "miracle" occurs in performing the sum over permutations, and one finds \cite{SasamotoHalfBrownReplica,SasamotoStationary,SasamotoStationary2}
that the overlap
takes the very simple "decoupled" form for the Brownian IC
\bea
&& \langle \Phi_0 | \Psi_\mu \rangle = n! 
\frac{ \prod_{j=1}^n (w_R + w_L - j) }{\prod_{j=1}^n (w_R - \frac{1}{2} - i \lambda_j) 
\prod_{j=1}^n (w_L - \frac{1}{2} + i \lambda_j) }
\eea
which leads to $\langle \Phi_0 | \Psi_\mu \rangle = n! 
\frac{1}{\prod_{j=1}^n (w_R - \frac{1}{2} - i \lambda_j)  }$ for the half-Brownian, and 
simply $\langle \Phi_0 | \Psi_\mu \rangle = n!$ for the droplet IC. In the infinite system
size limit, each eigenstate is made of $1 \leq n_s \leq n$ {\it strings} with 
rapidities $\lambda_{j,a}=k_j - \frac{i}{2} (m_j+1 - 2 a)$, $a=1,\dots,m_j$
(here $k_j$ are real momenta and $m_j \geq 1$ integers with $\sum_{j=1}^{n_s} m_j=n$).
The overlap are thus, for half-Brownian IC (hb) and Brownian IC (b)
\be
\langle \Phi_0 | \Psi_\mu \rangle_{hb} = n! \prod_{j=1}^{n_s}  s^{w_R}_{- k_j,m_j} 
\quad , \quad 
\langle \Phi_0 | \Psi_\mu \rangle_{b} = n! \frac{\Gamma(w_L + w_R)}{\Gamma(w_L + w_R - n)}
\prod_{j=1}^{n_s}  s^{w_R}_{- k_j,m_j}  s^{w_L}_{k_j,m_j} 
\quad , \quad s^w_{k,m}= \frac{\Gamma \left(w+i k-\frac{m}{2}\right)}{\Gamma
   \left(w+i k+\frac{m}{2}\right)}
\ee 
with $S^w_{k,1}= \frac{1}{w+ i k - \frac{1}{2}}$ and so on.
To remove the extra factor in the full Brownian case (and allow for a FD expression) one
defines (in that case only) the modified partition sum $\tilde Z(x,t)$ as
\be
\langle \tilde Z^n(x,t) \rangle =  \frac{ \langle Z^n(x,t) \rangle}{\prod_{j=1}^n (w_R + w_L - j)} \quad , \quad 
\tilde Z(x,t) = e^{\tilde h(x,t) } = e^{h(x,t) + \chi} \quad , \quad \langle e^{n \chi} \rangle 
= \frac{\Gamma(w_L + w_R - n)}{\Gamma(w_L + w_R)}
\ee 
i.e. one defines \cite{SasamotoStationary,SasamotoStationary2}
a randomly shifted height field $\tilde h(x,t) =h(x,t) + \chi$ (recalling that $Z(x,t)=e^{h(x,t)}$),
where $\chi$ a log-gamma variable of parameter $\gamma=w_L + w_R$, independent of $h$. 
Finally after inserting the known expression for the norms $||\mu||^2$ of the eigenstates, and
changing $k_j \to -k_j$ one obtains the formula \eqref{zbethe} of the text, which we reproduce
here 
\bea \label{zbethe2} 
&& \langle Z(x,t)^n \rangle = n! \sum_{n_s=1}^n \frac{1}{n_s!} \sum_{(m_1,\dots,m_{n_s})}  \prod_{j=1}^{n_s} \int \frac{dk_j}{2 \pi m_j}  e^{i x \sum_{j=1}^{n_s} m_j k_j} e^{t \sum_{j=1}^{n_s} [ \frac{1}{12} (m_j^3-m_j) - m_j k_j^2]}
\\
&& \times \prod_{j=1}^{n_s}  S_{k_j,m_j} 
\prod_{1 \leq i < j \leq n_s} \frac{4(k_i-k_j)^2 + (m_i-m_j)^2}{4(k_i-k_j)^2 + (m_i+m_j)^2} \nn
\eea 
with $S_{k,m}=1$ for the droplet IC, $S_{k,m}= s^w_{k,m}$ for the half-Brownian IC, and 
$S_{k,m}= s_{k,m}^{w_R} s_{-k,m}^{w_L}$ for the Brownian, where it is implicit 
here and below that in that case the l.h.s. of \eqref{zbethe2} must be replaced by $\langle \tilde Z(x,t)^n \rangle$,
the moments of the modified partition sum.

If the KP equation property holds, the cumulants must satisfy the equations \eqref{genN}. We want
to understand the mechanism for this property on the form \eqref{zbethe2}. Let us start with the
first two cumulants obtained from \eqref{zbethe2}. They read
\bea
&& Z_1(x,t)=\langle Z(x,t) \rangle = \int \frac{dk}{2 \pi}  e^{- i x k} e^{- t k^2}  S_{k,1}  \label{z1} \\
&& Z_2(x,t) = \langle Z(x,t)^2 \rangle - \langle Z(x,t) \rangle^2 =
e^{\frac{t}{2}} \int \frac{dk}{2 \pi}  e^{- 2 i x k} e^{- 2 t k^2} S_{k,2} \label{z2}  \\
&&  + \int \frac{dk_1}{2 \pi} \int \frac{dk_2}{2 \pi} 
e^{- i x (k_1+k_2)} e^{- t (k_1^2 + k_2^2) }
[ \frac{(k_1-k_2)^2}{(k_1-k_2)^2 + 1} -1 ] S_{k_1,1} S_{k_2,1} \nn
\eea 
In the expression for $Z_2$ the first integral is the contribution of the single string state which contains two bosons, $n_s=1$, $m_1=2$,
and the second integral the contribution of the two string state, $n_s=2$, $m_1=1$, $m_2=1$ (these
strings are just "particles" since their length is unity). \\

Let us now check that the first equation in \eqref{genN} is obeyed
\be \label{c2} 
\partial_t Z_2(x,t) 
- \frac{1}{2} Z_2(x,t) 
- \frac{1}{2} \partial_x^2 Z_2(x,t)  = 
\frac{1}{2} Z_{1}(x,t)^2 
\ee

We note that the differential operator $D_2=\partial_t - \frac{1}{2} \partial_x^2 - \frac{1}{2}$ gives zero on the first term 
in \eqref{z2}. It is the 1-string contribution $n_s=1$ and, as mentioned in the text, it is a
general property that this term obeys the {\it linear} part of the equation \eqref{c2} 
(and more generally the linear part of
\eqref{genN}). Acting on the
second term in \eqref{z2} the operator $D_2$ multiplies the integrand by 
\be
D_2(k_1,k_2)=- (k_1^2 + k_2^2) + \frac{1}{2} (k_1+k_2)^2 - \frac{1}{2} = - \frac{1}{2} ( (k_1-k_2)^2 +1) 
\ee
Hence in the integrand we have the simplification
\bea
D(k_1,k_2) [ \frac{(k_1-k_2)^2}{(k_1-k_2)^2 + 1} -1 ] =
D(k_1,k_2) \frac{-1}{(k_1-k_2)^2 + 1} = \frac{1}{2} 
\eea 
which leads to factorization of the double integral in two factors $Z_1$ given by
\eqref{z1}, which thus implies that \eqref{c2} holds. \\

%Let us now study $n=3$. One has
%\bea
%&& \langle Z(x,t)^3 \rangle = 2 e^{2 t} \int \frac{dk}{2 \pi}  e^{- 3 i x k} e^{- 3 t k^2} S_{k,3} 
%+ 3 e^{t/2} \int \frac{dk_1}{2 \pi} \frac{dk_2}{2 \pi}  e^{- i x (2 k_1 + k_2) - (2 k_1^2 + k_2^2) t} 
%S_{k_1,2} S_{k_2,1}  \frac{4(k_1-k_2)^2 + 1}{4(k_1-k_2)^2 + 9} \nonumber \\
%&&  + \int \frac{dk_1}{2 \pi} \frac{dk_2}{2 \pi} \frac{dk_3}{2 \pi}  
%e^{- i x (k_1 + k_2+k_3) - (k_1^2 + k_2^2+ k_3^2) t} S_{k_1,1} S_{k_2,1} S_{k_3,1} 
%\frac{(k_1-k_2)^2}{(k_1-k_2)^2 + 1}  \frac{(k_1-k_3)^2}{(k_1-k_3)^2 + 1}  \frac{(k_3-k_3)^2}{(k_2-k_3)^2 + 1} 
%\eea 

Let us check whether this mechanism, which uses only properties of the factor arising from
the norm, $\frac{(k_1-k_2)^2}{(k_1-k_2)^2}$, and not of the factor $S_{k,m}$, holds to
higher order. Let us write the third cumulant from \eqref{zbethe2}. We see that the 
substractions, arising from the definition of a cumulant, result in "counterterms" inside each
contribution, which we have written in a symmetric form
%
%and finally
%\bea
%&& \langle Z(x,t)^3 \rangle = 2 e^{2 t} \int \frac{dk}{2 \pi}  e^{- 3 i x k} e^{- 3 t k^2} S_{k,3} 
%+ 3 e^{t/2} \int \frac{dk_1}{2 \pi} \frac{dk_2}{2 \pi}  e^{- i x (2 k_1 + k_2) - (2 k_1^2 + k_2^2) t} 
%S_{k_1,2} S_{k_2,1}  \frac{4(k_1-k_2)^2 + 1}{4(k_1-k_2)^2 + 9} \nonumber \\
%&&  + \int \frac{dk_1}{2 \pi} \frac{dk_2}{2 \pi} \frac{dk_3}{2 \pi}  
%e^{- i x (k_1 + k_2+k_3) - (k_1^2 + k_2^2+ k_3^2) t} S_{k_1,1} S_{k_2,1} S_{k_3,1} 
%\frac{(k_1-k_2)^2}{(k_1-k_2)^2 + 1}  \frac{(k_1-k_3)^2}{(k_1-k_3)^2 + 1}  \frac{(k_3-k_3)^2}{(k_2-k_3)^2 + 1} 
%\eea 
\bea \label{3cum0} 
&& Z_3(x,t)=\langle Z(x,t)^3 \rangle- 3 \langle Z(x,t)^2 \rangle \langle Z(x,t) \rangle + 2 \langle Z(x,t) \rangle^3 
 = 2 e^{2 t} \int \frac{dk}{2 \pi}  e^{- 3 i x k} e^{- 3 t k^2} S_{k,3} \\
&& 
+ 3 e^{t/2} \int \frac{dk_1}{2 \pi} \frac{dk_2}{2 \pi}  e^{- i x (2 k_1 + k_2) - (2 k_1^2 + k_2^2) t} 
S_{k_1,2} S_{k_2,1} [ \frac{4(k_1-k_2)^2 + 1}{4(k_1-k_2)^2 + 9} - 1] \nonumber \\
&&  + \int \frac{dk_1}{2 \pi} \frac{dk_2}{2 \pi} \frac{dk_3}{2 \pi}  
e^{- i x (k_1 + k_2+k_3) - (k_1^2 + k_2^2+ k_3^2) t} S_{k_1,1} S_{k_2,1} S_{k_3,1}  \nn
\\
&& \times \bigg[ \frac{(k_1-k_2)^2}{(k_1-k_2)^2 + 1}  \frac{(k_1-k_3)^2}{(k_1-k_3)^2 + 1}  \frac{(k_3-k_3)^2}{(k_2-k_3)^2 + 1}
-  \frac{(k_1-k_2)^2}{(k_1-k_2)^2 + 1}  - \frac{(k_1-k_3)^2}{(k_1-k_3)^2 + 1}  - \frac{(k_2-k_3)^2}{(k_2-k_3)^2 + 1}  + 2
\bigg] \nn
\eea 
Again we have written first the contribution $n_s=1$, then $n_s=2$ and finally the $n_s=3$ term.
Let us check that the second equation from \eqref{genN} is obeyed. 
\be \label{c3} 
 \partial_t Z_3(x,t) 
- 2 Z_3(x,t) 
- \frac{1}{3} \partial_x^2 Z_3(x,t)  = 
4 Z_{1}(x,t) Z_{2}(x,t) 
\ee
The differential operator $D_3 = \partial_t - \frac{1}{3} \partial_x^2 - 2$, gives again zero when applied
on the first term $n_s=1$. 
On the second term $n_s=2$, it multiplies the integrand by
\be
D_3 \to - (2 k_1^2 + k_2^2) + \frac{1}{3} (2 k_1 + k_2)^2 - \frac{3}{2} = - \frac{1}{6} (4(k_1-k_2)^2 + 9) 
\ee 
thereby producing exactly the first term in the r.h.s of \eqref{c3} which reads explicitly
\bea \label{4zz} 
&& 4 Z_1(x,t) Z_2(x,t) = 4 e^{t/2} \int \frac{dk_1}{2 \pi} \frac{dk_2}{2 \pi}  e^{- i x (2 k_1 + k_2) - (2 k_1^2 + k_2^2) t} 
S_{k_1,2} S_{k_2,1} \\
&& + 4 
\int \frac{dk_1}{2 \pi} \frac{dk_2}{2 \pi} \frac{dk_3}{2 \pi}  
e^{- i x (k_1 + k_2+k_3) - (k_1^2 + k_2^2+ k_3^2) t} S_{k_1,1} S_{k_2,1} S_{k_3,1} 
[\frac{(k_1-k_2)^2}{(k_1-k_2)^2 + 1} -1] \nn
\eea 
Finally on the third term its action on the integrand gives exactly the second term in \eqref{4zz} using that
\bea
&& (-k_1^2 - k_2^2 - k_3^2 - \frac{1}{3} (k_1+k_2+k_3)^2 - 2) \\
&& \times \bigg[ \frac{(k_1-k_2)^2}{(k_1-k_2)^2 + 1}  \frac{(k_1-k_3)^2}{(k_1-k_3)^2 + 1}  \frac{(k_3-k_3)^2}{(k_2-k_3)^2 + 1}
-  \frac{(k_1-k_2)^2}{(k_1-k_2)^2 + 1}  - \frac{(k_1-k_3)^2}{(k_1-k_3)^2 + 1}  - \frac{(k_2-k_3)^2}{(k_2-k_3)^2 + 1}  + 2
\bigg] \nn \\
&& =  \frac{4}{3} [ \frac{(k_1-k_2)^2}{(k_1-k_2)^2 + 1} -1 + \text{2 perm}] \nn
\eea 
Hence \eqref{c3} is obeyed.\\

Although we have not established it $n \geq 4$, it is already clear on the cases $n=2,3$, that 
the mechanism of "simplification" which transforms the $n$-th cumulant onto a sum of lower cumulants,
upon application of the differential linear operator, works only from some combinatoric property of the norm factor,
and does not involve $S_{k,m}$. Only the decoupled form of the overlap is crucial, hence it
works {\it in exactly the same way} for droplet, half-Brownian, and Brownian (in the latter case using the 
modified partition sum $\tilde Z$). Of course this decoupled form is also the reason for
a simple FD formula to exist (when summing up the moments in the generating function $G$)
but it is useful to see how it works on the cumulants. \\

{\bf Remark 7.} One can ask how this mechanism works on the nested contour integral representation.
Consider any solution of the form 
\be
\langle Z(x,t)^n \rangle = \prod_j [\int_{C_j} \frac{dz_j}{2 i \pi} e^{t z_j^2 + x z_j} g(z_j)] 
\prod_{1 \leq i<j \leq n} \frac{z_i-z_j}{z_i-z_j-\bar c} 
\ee 
This formula holds for the droplet IC, with $\bar c=1$ in our units, where the $C_j$ are 
parallel to the imaginary axis with real parts such that ${\rm Re}(z_i-z_j) > \bar c$
(see formula (6.6) in Ref. \cite{Macdo}). 
In that case $g(z)=1$ but we consider here the more general case. Note that
the case $\bar c<0$ is also of interest as it provides a solution for the repulsive
delta Bose gas \cite{Coincidences}. \\

Let us set $\bar c=1$ (but the property extends for any $\bar c$).
For $n=2$ the property is very simple. The operator $D_2=\partial_t - \frac{1}{2} \partial_x^2 - \frac{1}{2}$ leads to
the multiplication of
the integrand by
\bea
z_1^2 + z_2^2 - \frac{1}{2} (z_1+z_2)^2 -\frac{1}{2}= \frac{1}{2} (z_1-z_2-1)(z_1-z_2+1) 
\eea 
The following simplification thus occurs in the integrand
\bea
\frac{1}{2} (z_1-z_2-1)(z_1-z_2+1)  (\frac{z_1-z_2}{z_1-z_2-1} -1) = \frac{1}{2} (z_1-z_2+1) \to \frac{1}{2}
\eea
The last step arises from the symmetry of the integrand, which can now be used, 
since the poles have disappeared and the contours $C_j$ can be brought together. Hence \eqref{c2} holds.
It works quite similarly to the previous paragraph, although the factors are slightly different. \\

For $n=3$, from the definition of the third cumulant in \eqref{3cum0}, we implement the subtractions 
in a symmetric way, which leads to the following factor in the integrand of $Z_3$ 
\bea \label{z3s} 
Z_3 \equiv \prod_{1 \leq i<j \leq n} \frac{z_i-z_j}{z_i-z_j-1} - \frac{z_1-z_2}{z_1-z_2-1} - \frac{z_1-z_3}{z_1-z_3-1} - \frac{z_2-z_3}{z_2-z_3-1} + 2 = \frac{2}{(z_1-z_2-1)(z_2-z_3-1)} 
\eea 
which, we note, simplifies. The differential linear operator $D_3$ acts by multiplying the 
integrand by $D_3(z_1,z_2,z_3) = z_1^2 + z_2^2 + z_3^2 - \frac{1}{3} (z_1+z_2+z_3)^2 - 2$
and an important property is that at the double pole of \eqref{z3s}, $D_3(z_1,z_2,z_3)|_{z_1=z_2+1,z_2=z_3+1} = 0$.
One also checks the following symmetrization property 
\bea
D_3(z_1,z_2,z_3) {\rm sym}_{z_1,z_2,z_3} [  \frac{2}{(z_1-z_2-1)(z_2-z_3-1)}]
= 4 {\rm sym}_{z_1,z_2,z_3} [  \frac{z_1-z_2}{z_1-z_2-1} - 1 ] 
\eea 
which is necessary condition for \eqref{c3} to hold. It would be sufficient for 
$\bar c<0$ but here for $\bar c>0$ one need to examine the poles to make
sure it holds also for the nested contours. Since we did that in the previous
paragraph, we know that it must work, and we will not pursue it here. It seems clear that 
there is a general mechanism for the equations on the cumulant to hold, 
provided, again, that the overlap is factorized.
It would be interesting to establish it for any value of $n$.

\section{Large time large deviation for more general $g(x)$, and linear statistics of the Airy process} 
\label{app:g} 

As noted in Remark 2., for any function $g(x)$, the function
$\phi(t,r)= \partial_r^2 q_{t,\beta}(\sigma=-e^{-r})$, where 
$q_{t,\beta}(\sigma=-e^{-r})$ is the FD \eqref{remark1}, must obey the KP equation
with a more general initial condition $\phi(t,r) \simeq_{t \to 0} \frac{1}{\sqrt{t}} p'_0(r)$
with 
\be
p'_0(r) = {\cal L}_1(-e^{-r})= \frac{\beta}{\pi} (\partial_r)^2 
\int_0^{+\infty} dx \sqrt{x} g(-e^{-x-r}) 
\ee
One can now choose a more general function $g(x)$, as in Eq. (214) of Ref. \cite{ShortTimeSystematic},
with
\be
\beta g(- e^{-r}) = \Gamma(1+\gamma) {\rm Li}_\gamma(- e^{- r})
\simeq_{r \to - \infty} - (- r)^{\gamma}_+
\ee 
except that we do not include the factor $t^{1-\gamma}$ of (214), i.e. we must choose $g(x)$ to be
time-independent. The value $\gamma=1$ corresponds to the KPZ case,
$g_{\rm KPZ}(x)=- \log(1-x)$. Then one has, from Eq. (218) of Ref. \cite{ShortTimeSystematic},
\be \label{asp0} 
p'_0(r) = {\cal L}_1(-e^{-r}) \simeq_{r \to - \infty}  - 
%\frac{\Gamma(\gamma+1)}{\sqrt{4 \pi} \Gamma(\gamma + \frac{1}{2})} 
\frac{\Omega}{2} 
(-r)^{\gamma- \frac{1}{2}}  \quad , \quad \Omega=\frac{\Gamma(1+\gamma)}{\sqrt{\pi} \Gamma(\frac{1}{2}+\gamma)}
\ee
which is a monomial at large negative $r$.\\

We now ask about the large time large deviation regime. We will use the analysis of cumulants
of Ref. \cite{ShortTimeSystematic}, extended here in Section \ref{subsec:cumulants}.
Let us perform the counting of powers of time. We write, from \eqref{scheme} and \eqref{res11} 
\bea
\log \hat G = \sum_{n \geq 1} \frac{\kappa_n(t,r)}{n!} = 
\sum_{n \geq 1} \frac{1}{n!} \bigg[ t^{\frac{n}{2}-1} 2^{n-1} (-1)^{n-1} (\partial_r)^{n-3} (p'_0(r))^{n} 
+ t^{\frac{n}{2}} (\partial_r)^{n-2}  (p'_0)^{n-2} + \dots \bigg]
\eea 
where the second (subleading) term is written only schematically. Because what we do is slightly
different than in Ref. \cite{ShortTimeSystematic} we must scale $r \sim t^\alpha$ and determine $\alpha$ later. We thus set $r=z t^\alpha$ with fixed $z<0$. Then the powers of time in the first
term are $t^{\frac{n}{2}-1 - \alpha (n-3) + n \alpha (\gamma- \frac{1}{2})}$. To make it independent of $n$ we must choose $\alpha=\frac{1}{3-2 \gamma}$ (we will restrict here to $\gamma<3/2$). 
The power of time of each term is then $\log \hat G \sim t^{3 \alpha-1} = t^{\frac{2 \gamma}{3 - 2 \gamma}}$.
The subleading term (second term) then scales as 
$\log \hat G \sim t^{\frac{n}{2} - \alpha (n-2) + (n-2) \alpha (\gamma- \frac{1}{2})}=t$, hence it is
indeed subdominant at large time for $\gamma > 3/4$, which is $\alpha>2/3$. Now we note that
although the power counting in $t$ is different, all coefficients being the same, the summation of the leading term should lead to the same function as in Ref. \cite{ShortTimeSystematic}, i.e 
\be \label{cumpred} 
\log \hat G(t,r) \simeq - t^{3 \alpha-1} \Phi_-(z) \quad , \quad \Phi_-(z)=- \frac{1}{2} \sum_{n \geq 1} 
\frac{(-\Omega)^n}{\Gamma(n+1)} \frac{\Gamma(n (\gamma- \frac{1}{2}) + 1)}{\Gamma(4 - n (\frac{3}{2}-\gamma))} (-z)^{3 - n (\frac{3}{2}-\gamma)} 
%\quad , \quad \Omega=\frac{\Gamma(1+\gamma)}{\sqrt{\pi} \Gamma(\frac{1}{2}+\gamma)}
\ee

On the other hand, we can directly search for a solution of the reduced KP equation
\eqref{KP2} which scales for large negative $r$ as 
\be
\log \hat G(t,r) = - t^{3 \alpha - 1} \Phi_-(\frac{r}{t^\alpha}) \quad , \quad 
\psi(t,r) = \partial_r^2 \log \hat G(t,r)  =  t^{\alpha-1} H_0(\frac{r}{t^\alpha})
\ee 
with $H_0=- \Phi_-''$. 
It leads to a generalisation of \eqref{eqH0}, to which it reduces for $\alpha=1$
\bea  \label{eqH0a} 
(\alpha-\frac{1}{2}) H_0(z) - \alpha z H_0'(z) + H_0(z) H_0'(z) = 0  
\eea 
This equation is solved by the change of variable $H_0(z)=z h(z)$ and $z=-e^u$ 
which leads to $\frac{dh}{du} = \frac{h(\frac{1}{2}-h)}{h-a}$ leading to
\be \label{use} 
- K z = (1 - 2 \frac{H_0(z)}{z})^{2 \alpha -1} (\frac{z}{H_0(z)})^{2 \alpha} 
\ee 
where $K$ is an integration constant. For $\alpha=1$ one recovers $H_0(z)=\frac{1}{K} (1- \sqrt{1- K z})$
with $K=\pi^2$ for the KPZ equation.

At large negative $z$, from \eqref{use} one has that $H_0(z) \simeq - K^{- \frac{1}{2 \alpha}} (-z)^{1 - \frac{1}{2 \alpha}}$. If we set $\alpha=\frac{1}{3-2 \gamma}$ as suggested by the above cumulant analysis, we
obtain $H_0(z) \simeq - K^{\gamma-\frac{3}{2}} (-z)^{\gamma- \frac{1}{2}}$. This is indeed the
behavior predicted in \eqref{cumpred} from the leading term $n=1$ (for $\gamma<3/2$ which we assume 
here), that is for $z \to -\infty$
\be
\Phi_-(z) \simeq \frac{\Gamma(1+\gamma)}{\sqrt{4 \pi} \Gamma(\frac{5}{2} + \gamma)} (-z)^{\gamma + \frac{3}{2}} \quad , \quad H_0(z) = - \Phi''_-(z) 
\simeq 
- \frac{\Omega}{2} 
%\frac{\Gamma(1+\gamma)}{\sqrt{4 \pi} \Gamma(\frac{1}{2} + \gamma)} 
(-z)^{\gamma - \frac{1}{2}} 
\ee 
hence we identify $K %= [ \frac{\Gamma(1+\gamma)}{\sqrt{4 \pi} \Gamma(\frac{1}{2} + \gamma)} ]^{\frac{-2}{3- 2 \gamma}}
= (\frac{\Omega}{2})^{\frac{-2}{3- 2 \gamma}}$ which reproduces $K=\pi^2$ for the KPZ case $\gamma=1$ (with $\Omega=2/\pi$). From \eqref{cumpred} 
the (large $|z|$) series expansion predicted by the cumulants reads
is 
\be
\frac{H_0(z)}{z} = - \frac{\Phi''_-(z)}{z} =  - \frac{1}{2} \sum_{n \geq 1} 
\frac{(-1)^n}{\Gamma(n+1)} \frac{\Gamma(n (\gamma- \frac{1}{2}) + 1)}{\Gamma(2 - n (\frac{3}{2}-\gamma))} (\Omega (-z)^{- (\frac{3}{2}-\gamma)})^n 
\ee 
which we can compare with the small $y,h= \frac{H_0}{z}$ expansion of the equation 
$y= h (1-2 h)^{\frac{1}{2 \alpha}-1}$ 
with $y=1/(-K z)^{1/(2 \alpha)}$. Setting $\alpha=\frac{1}{3-2 \gamma}$ one has
$y=\frac{\Omega}{2} (-z)^{- (\frac{3}{2}-\gamma)}$ and the equation becomes
$y= h (1-2 h)^{\frac{1}{2}-\gamma}$. It is then easy to check with Mathematica
that the two series coincide.\\

Hence the cumulant method and the KP equation once again agree, now for a larger class of functions $g$, i.e. a larger class of linear statistics of the Airy$_2$ point process.
One recovers then from the KP equation the results for the large deviation function $\Phi_-(z)$
of Ref. \cite{ShortTimeSystematic} and Ref. \cite{largedev} for monomials $x_+^\gamma$, although in a slightly different setting.\\

Let us close by showing how the above results can be equivalently be derived using the semi-classical expansion discussed in Section \ref{subsec:cumulants}, which, to leading order, maps to the Burgers equation. We can use the solution
of the Burgers equation \eqref{solub}, in the form $\psi_0(t,r) = \frac{1}{\sqrt{t}} p_0'(r - 2 t \psi_0(t,r))$.
In the large time limit we insert the scaling form $\psi_0(t,r) = t^{\alpha-1} H_0(r/t^\alpha)$
and use the asymptotics \eqref{asp0} of $p_0'(r)$ at large negative $r= z t^\alpha$. One sees that the powers of $t$ cancel out only if $\alpha=1/(3-2 \gamma)$, in agreement with the above results, and we obtain
\be
H_0 = - \frac{\Omega}{2} (-z + 2 H_0)^{\gamma-\frac{1}{2}} 
\ee 
It is easy to see that this equation is equivalent to the equation \eqref{use} with $K= (\frac{\Omega}{2})^{\frac{-2}{3- 2 \gamma}}$ in full agreement with the above results. This already indicates that the Burgers equation
solution is equivalent to the self-consistent equation (15) found in Ref. \cite{largedev}, but we leave
the full analysis of these connections to a future work.

\section{Cumulants for the half-Brownian IC in the small time large deviation regime}
\label{app:cum} 

We give explicit expressions for the cumulants at short time for the half-Brownian IC. 
One can solve directly the linear equations for the cumulants 
in the small time large deviation regime, from the main text
\bea \label{genN4} 
 [ \partial_t - \frac{n^3-n}{12} -  \frac{1}{n} \partial_x^2 ] Z_n(x,t)  = 0
\eea 
One can look for solutions of the form
\be
Z_n(x,t) \simeq \frac{1}{\sqrt{\pi t}} t^{n/2} F_n(y= x/\sqrt{t}) 
\ee
Inserting we find that it is 
%\be
%Z_n(x,t)  = n^{n-2} x^{n-1} e^{-n w} p_n(x/\sqrt{t})   \quad , \quad p_n(+\infty)=1  \quad , \quad p_n(-\infty)=0
%\ee 
%This leads to the equation for $p_n$, for $w=0$
%\be
%2 \left(n^2-3 n+2\right) p(y)+y \left(\left(n
%   \left(y^2+4\right)-4\right) p'(y)+2 y p''(y)\right) = 0
%\ee
%This is 
solved by hypergeometric functions. One finds, for $w=0$
\bea
&& F_n(y) = 2^{n-2} n^{\frac{n-3}{2}} \left(\Gamma
   \left(\frac{n}{2}\right) \,
   _1F_1\left(\frac{1-n}{2};\frac{1}{2};-\frac{n
   y^2}{4}\right)+\sqrt{n} y \Gamma
   \left(\frac{n+1}{2}\right) \,
   _1F_1\left(1-\frac{n}{2};\frac{3}{2};-\frac{n
   y^2}{4}\right)\right)
\eea
leading to the explicit forms for e.g. the lowest cumulants 
\bea
&& Z_1(x,t) \simeq \frac{1}{2} \left(\text{erf}\left(\frac{x}{2
   \sqrt{t}}\right)+1\right) \\
   && Z_2(x,t) \simeq \frac{1}{2} x \left(\text{erf}\left(\frac{x}{\sqrt{2}
   \sqrt{t}}\right)+1\right)+\frac{\sqrt{t}
   e^{-\frac{x^2}{2 t}}}{\sqrt{2 \pi }} \\
   && Z_3(x,t) \simeq \frac{1}{2} \left(2 t+3 x^2\right)
   \left(\text{erf}\left(\frac{\sqrt{3} x}{2
   \sqrt{t}}\right)+1\right)+\sqrt{\frac{3}{\pi }}
   \sqrt{t} x e^{-\frac{3 x^2}{4 t}}
\eea
and one can check that they have the correct $t=0$ limits
\eqref{incond}. For $w>0$ one finds
\bea
&& Z_n(x,t) \simeq \frac{1}{\sqrt{\pi t}} t^{n/2} F_n(y= x/\sqrt{t}) \\
&& F_n(y) = 
2^{n-2} n^{\frac{n-3}{2}} e^{-\frac{n y^2}{4}}
   \left(\Gamma \left(\frac{n}{2}\right) \,
   _1F_1\left(\frac{n}{2};\frac{1}{2};\frac{1}{4} n
   (y-2 {\tilde w})^2\right)+\sqrt{n} \Gamma
   \left(\frac{n+1}{2}\right) (y-2 {\tilde w}) \,
   _1F_1\left(\frac{n+1}{2};\frac{3}{2};\frac{1}{4} n
   (y-2 {\tilde w})^2\right)\right) \nn
\eea

\section{Fredholm determinant and KP equation}
\label{app:fd} 

In a seminal paper \cite{Poppe1988,Poppe1989} P\"oppe and  Sattinger found a family of Fredholm determinants (FD) which satisfy the KP hierarchy. Following their (redundant) notation, consider $X=(x_1,x_2,x_3,..)$, $Z=(z_1,x_2,x_3,..)$ and a kernel $F(X,Z)$ which 
satisfies the linear equation (Eq. (2.4) in Ref. \cite{Poppe1988})
\bea \label{poppe1} 
\partial_{x_n} F - \partial_{x_1}^n F + (-)^n \partial_{z_1}^n F = 0 \quad, \quad n=2,3.. 
\eea 
Note that $x_n$ for $n>1$ occurs in both $X$ and $Z$, hence $\partial_{x_n}$
acts on both arguments of the kernel. One then defines the FD noted $D(X)$ on $\mathbb{L}^2[x_1,+\infty[$, with $x_2,x_3,\dots$ being parameters
\be
D(X) = {\rm Det}(I + P_{[x_1,+\infty[} F) = \sum_{n=0}^{+\infty} \frac{1}{n!} 
\prod_{a=1}^n \int_{x_1}^{+\infty} dy_1^a \det_{1 \leq b,c  \leq n} F(Y_{b},Y_{c}) 
\quad , \quad Y_a=(y^a_1,x_2,x_3,\dots) 
\ee 
and $P_{[x_1,+\infty[}$ the projector on the interval $[x_1,+\infty[$. 
In other words $D(X)$ is the standard FD on $\mathbb{L}^2[x_1,+\infty[$ for the operator $\tilde F$
with kernel $\tilde F(y,y')=F((y,x_2,\dots),(y',x_2,\dots))$. Then (Theorem 3.1 in \cite{Poppe1988}) the function
\bea
u(X)=2 \partial_{x_1}^2 \log D(X)
\eea 
satisfies the KP hierarchy, $D(X)$ being the tau function. The lowest member is the KP equation, obtained for $X=(x_1,x_2,x_3)$,
which reads (in the conventions of Ref. \cite{Poppe1988})
\bea
\partial_{x_3} u - \frac{1}{4} (\partial_{x_1}^3 u + 6 u \partial_{x_1} u) = \frac{3}{4} \partial_{x_1}^{-1} \partial_{x_2}^2 u 
\eea 

To connect with the present paper we set 
\be
x_1 = r \quad , \quad x_2 = \frac{x}{2} \quad , \quad x_3 = - \frac{t}{3} \quad , \quad u(x_1,x_2,x_3) 
= 2 \phi(x,t,r) \quad , \quad D(x_1,x_2,x_3)  = G(x,t,r) 
\ee 
Let us now define the following kernel on 
$\mathbb{L}^2[0,+\infty[$ ($v,v'>0$) 
\be
K_{xtr}(v,v')= K_{xt}(r+ v,r + v') = F((v+r,\frac{x}{2},- \frac{t}{3}),(v'+r,\frac{x}{2},- \frac{t}{3}))
\ee 
where $x,t$ are parameters. By construction it satisfies
\be \label{co1} 
 \partial_r K_{xtr}(v,v') = (\partial_v + \partial_{v'}) K_{xtr}(v,v') 
\ee
and the conditions obtained from \eqref{poppe1} read
\bea \label{co2} 
 \partial_t K = - \frac{1}{3} (\partial_v^3 + \partial_{v'}^3) K \quad , \quad  \partial_x K = \frac{1}{2} (\partial_v^2 - \partial_{v'}^2) K
\eea 
Hence, if these conditions are satisfied, one has that 
\be
\phi(x,t,r) = \partial_r^2 {\rm Det}(1 + \alpha P_{[0,+\infty[} K_{xtr})
\ee
satisfies the KP equation \eqref{KP1} for any $\alpha$ such that the FD is well-defined. 
Eqs \eqref{co1},\eqref{co2} are the conditions given in Ref. \cite{QRKP}.

The generating function for the droplet IC can be written as
\bea
&& G(x,t,r)= {\rm Det}(I - M_{xtr})|_{\mathbb{L}^2(\mathbb{R}^+)} \\
&& M_{xtr}(v,v') = \int du \, \Sigma(t^{1/3} u - r) \, {\rm Ai}(u+v+ \frac{x^2}{4 t^{4/3}}) {\rm Ai}(u+v'+ \frac{x^2}{4 t^{4/3}})
 \quad , \quad \Sigma(z)= \frac{1}{1 + e^{-z}} 
\eea 
Performing the shift $u \to u + t^{-1/3} r$, 
using the integral representation of both Airy functions, rescaling $z,w \to t^{1/3} z, t^{1/3} w$, $u \to t^{-1/3} u$, and, in a second stage, using the translation $z \to z + \frac{x}{2 t}$, $w \to w - \frac{x}{2 t}$ we see that 
$M$ is equivalent under a similarity transformation (which does not change the FD) to the kernel
\be \label{12} 
K_{xtr}(v,v') = \int_{-\infty}^{+\infty} du \Sigma(u) \int_{C^2} \frac{dz dw}{(2 i \pi)^2} e^{ t (\frac{z^3}{3} + \frac{w^3}{3}) - z (v + r + u) 
- w (v'+r+u) + x (\frac{z^2}{2} - \frac{w^2}{2}) }
\ee  
with $M_{xtr}(v,v') = t^{1/3} K_{xtr}(v t^{1/3}, v' t^{1/3}) e^{\frac{x}{2 t} (v'-v)}$. The kernel $K$ manifestly satisfies
the above conditions Eqs \eqref{co1},\eqref{co2}. This establishes that for the droplet IC,
$\phi(x,t,r) = \partial_r^2 G(x,t,r)$ satisfies the KP equation.\\

Furthermore, the same conditions Eqs \eqref{co1},\eqref{co2} are also satisfied for any choice of $\Sigma(u)$ in \eqref{12} (for which we assume the FD to be well defined).
For the choice $\Sigma(u)=1 - e^{\beta g(- e^u)}$ one recovers exactly the FD considered in the
Remark 2 (and in Ref. \cite{ShortTimeSystematic}). Indeed one then has
${\rm Det}(I - M_{xtr})|_{\mathbb{L}^2(\mathbb{R}^+)}={\rm Det}[ I - (1- e^{\beta \hat g_{t,\sigma}}) K_{\rm Ai}]$
where $\hat g_{t,\sigma}(u)=g(\sigma e^{t^{1/3} u})$ and $\sigma=-e^{-r}$, as in \eqref{remark1}
(this is easily seen e.g. expanding in traces and using the cyclic property). Hence
the whole class of FD \eqref{remark1}, useful to evaluate linear statistics of the
Airy point process, satisfy the KP property, as claimed in the text.\\

Finally note that the $N$ soliton solution of the KP hierarchy is obtained from a linear superposition of a
particular solution of \eqref{poppe1}
\be
F = \sum_{j=1}^N a_j e^{ x_1 p_j - z_1 q_j +  \sum_{n \geq 2} x_n (p_j^n - q_j^n)}
\ee 
One finds that 
\be
D(x) = \det_{N \times N} \bigg( \delta_{ij} - \frac{a_i}{p_i-q_j} e^{ x_1 (p_i - q_j) 
+  \sum_{n \geq 2} x_n (p_i^n - q_j^n)} \bigg)
\ee 
More general solutions are obtained from the continuous superposition
\be
F = \int d\mu(p,q) e^{ x_1 p - z_1 q +  \sum_{n \geq 2} x_n (p^n - q^n)}
\ee 
for some weight measure $d \mu(p,q)$, leading to kernels generalizing \eqref{12}
and which obey the KP hierarchy property.


\begin{thebibliography}{12}
\bibitem{footnoteQR}
Note that for the half-Brownian this was noted in the second version of the work 
\cite{QRKP}.
\end{thebibliography}
\end{document}